\documentclass[11pt,a4paper]{article}
\pdfoutput=1
\usepackage[pdftex]{graphics}
\usepackage{jheppub}
\usepackage{amsmath,amssymb,amsfonts}
\usepackage{enumitem}
\usepackage{multirow}
\usepackage{array,booktabs}
\usepackage{slashed}

\usepackage{accents}
\usepackage{mathrsfs}
\usepackage{mathtools}

\usepackage[usenames,dvipsnames]{xcolor}
\definecolor{labelcolor}{RGB}{194, 175, 116}
\definecolor{rmkcolor}{RGB}{215,30,5}
\definecolor{feyntext}{RGB}{20,125,233}
\definecolor{lred}{RGB}{255,130,130}
\definecolor{llred}{RGB}{255,160,160}
\definecolor{skyblue}{RGB}{34,139,230}
\definecolor{navy}{rgb}{0,0,0.7}
\definecolor{purple}{RGB}{171,1,207}
\definecolor{lgreen}{RGB}{231, 242, 43}
\definecolor{lgray}{RGB}{135, 135, 145}
\definecolor{recur}{RGB}{231, 242, 43}
\usepackage[export]{adjustbox}
\usepackage[final]{showlabels} %

\newcommand{\Fig}[1]{Fig.\,\ref{#1}}
\renewcommand{\eqref}[1]{Eq.\,(\ref{#1})}
\newcommand{\eqrefs}[2]{Eqs.\,(\ref{#1}) and (\ref{#2})}
\newcommand{\Sec}[1]{Sec.\,\ref{#1}}

\newcommand{\App}[1]{App.\,\ref{#1}}
\newcommand{\rcite}[1]{Ref.\,\cite{#1}}
\newcommand{\rrcite}[1]{Refs.\,\cite{#1}}
\def\mem{\hspace{0.1em}}
\def\hem{\hspace{0.05em}}
\def\nem{\hspace{-0.1em}}
\def\hnem{\hspace{-0.05em}}
\def\hhem{\hspace{0.025em}}

\def\blank{{\,\,\,\,\,}}

\def\qiq{{\quad\implies\quad}}
\def\qfq{{\quad\iff\quad}}

\def\a{\alpha}

\def\e{\epsilon}
\def\ve{\varepsilon}

\def\s{\sigma}

\def\L{\Lambda}
\def\bpsi{{\smash{\bar{\psi}}\kern0.02em\vphantom{\psi}}}
 \def\mathe{{\scalebox{1.01}[1]{$\mathrm{e}$}}}

\def\mtimes{{\mem\times\mem}}
\def\mdot{{\mem\cdot\mem}}

\def\tensor\otimes

\def\MPl{{M_\text{Pl}}}

\def\lsq{{
		\kern-0.037em
		\adjustbox{scale=0.90,valign=c}{$
			{
				\adjustbox{raise=-0.09em}{$\lfloor$}
				\llap{\reflectbox{\rotatebox[origin=c]{180}{$\lfloor$}}}
			}
			$}
		\kern-0.04em
}}
\def\rsq{{
		\kern-0.04em
		\adjustbox{scale=0.90,valign=c}{$
			{
				\rlap{\reflectbox{\rotatebox[origin=c]{180}{$\rfloor$}}} 
				\adjustbox{raise=-0.09em}{$\rfloor$}
			}
			$}
		\kern-0.037em
}}

\def\O{\mathcal{O}}

\def\G{\mathcal{G}}

\newcommand{\BB}[1]{\Big(\,{#1}\,\Big)}
\newcommand{\bb}[1]{\bigg(\,{#1}\,\bigg)}

\newcommand{\bbsq}[1]{\bigg[\,{#1}\,\bigg]}

\newcommand{\bigbig}[1]{\big(\mem{#1}\mem\big)}

\def\P{\mathcal{P}}

\newcommand{\dbar}{
	d\kern-.20em\makebox[0pt][l]{\kern0.01em\adjustbox{raise=-0.01em}{\scalebox{1.4}[1.0]{$\bar{}$}}\kern-0.01em}\kern.20em
}
\newcommand{\deltabar}{
	\delta\kern-.20em\makebox[0pt][l]{\adjustbox{raise=-0.01em}{\scalebox{1.0}[1.0]{$\bar{}$}}}\kern.20em
}

\usepackage{bm}

\usepackage{dsfont}

\def\M{{\mathcal{M}}}

\newcommand{\expval}[1]{
	\big\langle\hem{
		#1
	}\hem\big\rangle
}

\let\oldcap\cap
\renewcommand{\cap}{{\,\oldcap\,}}

\setlist[itemize]{
	label=\adjustbox{scale=0.7}{$\bullet$}, itemsep=0pt,topsep=0px
}
\setlist[enumerate]{
	itemsep=0pt,topsep=3px
}

\usepackage{tikz}
\usetikzlibrary{calc} %
\usetikzlibrary{shapes.geometric} %
\usetikzlibrary{positioning} %
\usetikzlibrary{fit} %
\usepackage[a]{esvect} %
\usetikzlibrary{decorations.pathmorphing}
\tikzset{every node/.style = {inner sep = 0pt, outer sep = 0, minimum size = 0}}
\tikzset{t/.style = {
		inner sep = 1.5pt, outer sep =1.5pt, minimum size = 1pt,
		font = \small, text = feyntext
}}
\tikzset{T/.style = {
		inner sep = 1.5pt, outer sep =1.5pt, minimum size = 1pt,
		font = \small
}}

\tikzset{linear/.style = {draw, line width = 1.2pt}}
\tikzset{wiggly/.style = {draw, line width = 1.2pt,
		decorate, decoration={snake, 
			amplitude=1.5pt, segment length=5.0pt, post length=0pt, pre length=0pt
		}
}}
\tikzset{sdot/.style = {circle, fill=black, inner sep=0pt, outer sep=0pt, minimum size=1.0pt}}
\tikzset{tdot/.style = {circle, fill=black, inner sep=0pt, outer sep=0pt, minimum size=1.5pt}}
\tikzset{cdot/.style = {circle, draw=black, fill=black, inner sep=0pt, outer sep=0pt, minimum size=2.0pt, line width=1.2pt}}
\tikzset{odot/.style = {circle, draw=black, fill=lred, inner sep=0pt, outer sep=0pt, minimum size=4.0pt, line width=1.2pt}}

\tikzset{gdot/.style = {circle, draw=black, fill=lgray, inner sep=0pt, outer sep=0pt, minimum size=4.0pt, line width=1.2pt}}

\tikzset{wdot/.style = {circle, draw=black, fill=white, inner sep=0pt, outer sep=0pt, minimum size=6.0pt, line width=1.2pt}}
\tikzset{bdot/.style = {
		circle, fill=black, inner sep=0pt, outer sep=0pt, minimum size=6.0pt, line width=1.2pt,
		opacity=0.8,
		postaction={
			circle, draw=black, inner sep=0pt, outer sep=0pt, minimum size=6.0pt, line width=1.2pt
			,pattern={mylines[size=1.5pt, line width=0.8pt, angle=45]},
			pattern color=white,
			opacity=1.0,
		}
}}

\tikzset{poly2/.style={
		append after command={\pgfextra{
				\filldraw[fill=lgray, draw=black, line width=1.2pt]
				($(\tikzlastnode.center) - (0,0.344146em)$) arc[
				start angle=-35, end angle=35, radius= 0.6em
				] arc[
				start angle=-35, end angle=35, radius=-0.6em
				] -- cycle;
		}}
}}
\tikzset{poly1/.style={
		append after command={\pgfextra{
				\filldraw[fill=lgray, draw=black, line width=1.2pt]
				($(\tikzlastnode.center)$)
				--
				($(\tikzlastnode.center) + ( 45:0.0722)$)
				arc[
				start angle=135, end angle=-135, radius=0.0722
				]
				-- 
				($(\tikzlastnode.center) + (-45:0.0722)$)
				-- cycle;
		}}
}}

\tikzset{poly3/.style = {
		draw,
		regular polygon, 
		regular polygon sides = 3,
		line width = 1.2pt,
		fill = lgray,
		minimum size = 5.8pt,
		rotate=30
}}
\tikzset{poly4/.style = {
		draw,
		regular polygon, 
		regular polygon sides = 4,
		line width = 1.2pt,
		fill = lgray,
		minimum size = 5.8pt
}}

\tikzset{yblob/.style = {
regular polygon,
regular polygon sides = 4,
draw=black, fill=llred, inner sep=0pt, outer sep=0pt, minimum size=15pt, line width=1.2pt,
font = \footnotesize
}}
\tikzset{squ/.style = {
		regular polygon,
		regular polygon sides = 4,
		draw=black, fill=white, inner sep=0pt, outer sep=0pt, minimum size=12pt, line width=1.2pt
}}
\tikzset{zsqu/.style = {
		regular polygon,
		regular polygon sides = 4,
		draw=black, fill=white, inner sep=0pt, outer sep=0pt, minimum size=14pt, line width=1.2pt,
		font = \footnotesize
}}
\tikzset{ysqu/.style = {
		regular polygon,
		regular polygon sides = 4,
		draw=black, fill=recur, inner sep=0pt, outer sep=0pt, minimum size=14pt, line width=1.2pt,
		font = \footnotesize
}}

\usetikzlibrary{patterns}
\usetikzlibrary{patterns.meta}
\tikzdeclarepattern{
	name=mylines,
	parameters={
		\pgfkeysvalueof{/pgf/pattern keys/size},
		\pgfkeysvalueof{/pgf/pattern keys/angle},
		\pgfkeysvalueof{/pgf/pattern keys/line width},
	},
	bounding box={
		(0,-0.5*\pgfkeysvalueof{/pgf/pattern keys/line width}) and
		(\pgfkeysvalueof{/pgf/pattern keys/size},
		0.5*\pgfkeysvalueof{/pgf/pattern keys/line width})},
	tile size={(\pgfkeysvalueof{/pgf/pattern keys/size},
		\pgfkeysvalueof{/pgf/pattern keys/size})},
	tile transformation={rotate=\pgfkeysvalueof{/pgf/pattern keys/angle}},
	defaults={
		size/.initial=5pt,
		angle/.initial=45,
		line width/.initial=.4pt,
	},
	code={
		\draw [line width=\pgfkeysvalueof{/pgf/pattern keys/line width}]
		(0,0) -- (\pgfkeysvalueof{/pgf/pattern keys/size},0);
	},
}
\tikzset{arb-blob/.style = {
		circle, fill=lgray, inner sep=0pt, outer sep=0pt, minimum size=10pt, line width=0pt,
		postaction={
			circle, draw=black, inner sep=0pt, outer sep=0pt, minimum size=8pt, line width=1.2pt
			,pattern={mylines[size=1.5pt, line width=0.8pt, angle=45]},
			pattern color=white
		}
}}
\tikzset{arb-blob-red/.style = {
		circle, fill=lred, inner sep=0pt, outer sep=0pt, minimum size=10pt, line width=0pt,
		postaction={
			circle, draw=black, inner sep=0pt, outer sep=0pt, minimum size=8pt, line width=1.2pt
			,pattern={mylines[size=1.5pt, line width=0.8pt, angle=45]},
			pattern color=white
		}
}}

\tikzset{gray-blob/.style = {
		circle, fill=lgray, inner sep=0pt, outer sep=0pt, minimum size=8pt, draw, line width=1.2pt
}}

\usetikzlibrary{arrows, decorations.markings}
\makeatletter
\def\pgf@lib@dec@parsenum#1{%
	\gdef\pgf@lib@dec@computed@width{0 pt}%
	\tsx@pgf@lib@dec@parsenum#1+endmarker+%
	\ifdim\pgf@lib@dec@computed@width<0pt\relax%
	\pgfmathparse{\pgfdecoratedpathlength\pgf@lib@dec@computed@width}
	\edef\pgf@lib@dec@computed@width{\pgfmathresult pt}%
	\fi%
}

\def\tsx@pgf@lib@dec@parsenum@endmarker{endmarker}

\def\tsx@pgf@lib@dec@parsenum#1+{
	\def\temp{#1}%
	\ifx\temp\tsx@pgf@lib@dec@parsenum@endmarker%
	\else%
	\tsx@pgf@lib@dec@parsenum@one{#1}%
	\expandafter\tsx@pgf@lib@dec@parsenum%
	\fi%
}

\def\tsx@pgf@lib@dec@parsenum@one#1{%
	\pgfmathparse{#1}%
	\ifpgfmathunitsdeclared%
	\pgfmathparse{\pgf@lib@dec@computed@width + \pgfmathresult pt}%
	\else%
	\pgfmathparse{\pgf@lib@dec@computed@width + \pgfmathresult*\pgfdecoratedpathlength*1pt}%
	\fi%
	\edef\pgf@lib@dec@computed@width{\pgfmathresult pt}%
}
\makeatother

\usetikzlibrary{arrows.meta}
\tikzset{
	lin/.style = {
		draw, line width=1.2pt
	}
}
\tikzset{
	prop/.style = {
		draw, line width=1.2pt, 
		decoration = { 
			markings, 
			mark = at position 0.5 + 3.2pt with {
				\arrow{>[length=5pt,width=5pt]}
			}
		},
		postaction = {decorate}
	}
}
\tikzset{
	dprop/.style = {
		draw, line width=1.2pt,
		dotted, 
		line cap=round,
		decoration = { 
			markings, 
			mark = at position 0.5 + 3.2pt with {
				\arrow{>[length=5pt,width=5pt]}
			}
		},
		postaction = {decorate}
	}
}
\tikzset{arrp/.style={-{Circle  [length=3.2pt,width=3.2pt]}}}
\tikzset{arrv/.style={-{Circle  [length=4pt,width=4pt, open]}}}

\tikzset{
	proper/.style = {
		draw, line width=1.2pt, 
		decoration = { 
			markings, 
			mark = at position 0.5 + 3.2pt with {
				\arrow{>[length=5pt,width=5pt]}
			}
		},
		postaction = {decorate}
	}
}

\tikzset{
	fprop/.style = {
		draw, line width=1.2pt, 
		decoration = { 
			markings, 
			mark = at position 0.5 + 5.4pt with {
				\arrow{Triangle[length=10pt,width=6pt,fill=lgray]}
			}
		},
		postaction = {decorate}
	}
}
\tikzset{
	qprop/.style = {
		draw, line width=1.2pt, 
		decoration = { 
			markings, 
			mark = at position 0.5 + 3.95pt with {
				\arrow{Triangle[length=6pt,width=3.6pt]}
			}
		},
		postaction = {decorate}
	}
}
\tikzset{wprop/.style = {
		draw, line width = 1.2pt,
		line cap = round,
		line join = round,
		decorate, decoration={
			zigzag,
			amplitude=0.8pt,
			segment length=3.0pt, 
			post length=0pt, pre length=0pt
		}
}}

\tikzset{empty/.style = {inner sep = 0pt, outer sep = 0, minimum size = 0}}
\tikzset{w/.style = {inner sep = 1pt, outer sep = 2pt, minimum size = 12pt, anchor = west,	
	font = \small
}}

\newcommand{\kk}[1]{{k{\kern-0.17em}k_{#1}}}
\newcommand{\kf}[1]{{k{\kern-0.17em}f_{#1}}}
\newcommand{\ff}[1]{{f{\kern-0.27em}f_{#1}}}

\def\R{\mathbb{R}}

\def\diff{\mathfrak{diff}}

\def\GL{\mathrm{GL}}

\def\tf{{\smash{\tilde{f}}}{}}

\def\symp{\mathrm{symp}}
\def\Symp{\mathrm{Symp}}
\def\diff{\mathrm{diff}}
\def\Diff{\mathrm{Diff}}

\definecolor{if}{RGB}{84,85,109}

\definecolor{i}{RGB}{210,0,0}
\definecolor{f}{RGB}{0,154,0}

\definecolor{i}{RGB}{242,63,28}
\definecolor{f}{RGB}{150,82,232}

\def\ini{{\texttt{i}}}
\def\fini{{\texttt{f}}}

\def\ti{{t_\ini}}
\def\tf{{t_\fini}}
\def\tt{(t_\fini{\mem-\mem\hem}t_\ini)}

\def\L{{\mathcal{L}}}

\newcommand{\Pexp}[1]{
    \mathrm{T}\kern-0.1em\exp\nem
    \bigg(\hem{
        #1
    }\bigg)
}

\def\lb{\{\kern-0.15em\{}
\def\rb{\}\kern-0.15em\}}

\newcommand{\pb}[2]{{\{\hem{#1},{#2}\hem\}}}

\definecolor{skyblue}{RGB}{34,139,230}
\definecolor{indigo}{RGB}{63,72,204}
\definecolor{olg}{RGB}{100,154,0}
\definecolor{or}{RGB}{253,111,0}
\definecolor{pink}{RGB}{255, 69, 149}

\definecolor{emphcolor}{RGB}{207,0,18}

\usepackage[outline]{contour}
\contourlength{0.1pt}

\renewcommand{\emph}[1]{\textcolor{emphcolor}{\contour{emphcolor}{{\textit{#1}}}{\kern0.07em}}}

\newcommand{\defn}[1]{\textcolor{defcolor}{\contour{defcolor}{{\textit{#1}}}{\kern0.07em}}}

\def\G{\Omega}

\newcounter{alphnum}

\counterwithin*{alphnum}{subsubsection}

\def\Del{\mathit{\Delta}}

\DeclareMathOperator{\Ext}{Ext}

\def\trho{\tilde{\rho}}
\def\tH{\tilde{H}}

\newcommand{\Extr}[2]{
	\underset{\substack{
		#1
	}}{\Ext}\,
	\bbsq{
		#2
	}
}

\def\ba{{\bar{a}}}
\def\C{\mathbb{C}}

\title{
	Manifest Symplecticity
	in
	Classical Scattering
}

\author[a]{Joon-Hwi Kim}

\affiliation[a]{Walter Burke Institute for Theoretical Physics,\\ California Institute of Technology, Pasadena, CA 91125}

\abstract{
	The Liouville theorem states that
	classical time evolution is an incompressible flow in phase space.
	We investigate two formulations
	of classical mechanics 
	in which this property is manifested.
	First, the traditional Hamilton-Jacobi theory
	provides an in-out formalism.
	Second, a recent idea
	employing an exponential representation of time evolution
	provides an in-in formalism.
	Through explicit examples,
	it is demonstrated that
	the on-shell action in the former
	and
	the exponential generator in the latter
	are disparate objects.
	Still, a concrete relation between the two is identified
	in terms of a matching calculation.
	A strictly classical derivation and formulation of
	classical scattering theory is provided.
}

\renewcommand*{\bibfont}{\footnotesize}

\begin{document}

\begin{flushright}
    \footnotesize
    CALT-TH 2025-035
\end{flushright}
\maketitle

\bibliographystyle{utphys-modified}
\renewcommand*{\bibfont}{\footnotesize}

\section{Introduction}

Picture two giant black holes orbiting each other,
creating gravitational waves
that will eventually reach our detectors at LIGO
\cite{LIGOScientific:2016aoc,LIGOScientific:2017vwq}.
To make theoretical predictions for
such gravitational wave signals,
it is necessary to grasp the classical gravitational dynamics of macroscopic astrophysical bodies.
Interestingly, modern physicists find themselves using
quantum field theory, i.e.,
the very framework developed for microscopic particles,
to efficiently perform high-precision computations
for such theoretical predictions
\cite{KMOC,%
Bern:2021dqo,Bern:2019crd,%
Bern:2019nnu,Cheung:2018wkq,%
Neill:2013wsa,%
Levi:2018nxp,Porto:2016pyg,Goldberger:2004jt,%
Kalin:2020mvi,Kalin:2020fhe,%
Mogull:2020sak,Jakobsen:2021zvh,Jakobsen:2021smu,Jakobsen:2022psy,Kalin:2022hph,Jakobsen:2023ndj,Jakobsen:2023pvx,%
Dlapa:2021vgp,Dlapa:2022lmu,Dlapa:2023hsl,
Driesse:2024xad,%
Dlapa:2024cje,Dlapa:2025biy,%
Haddad:2025cmw,He:2025how%
}.
The idea is to think of astrophysical objects
effectively as point particles
and study their scattering amplitudes,
encoding physical information 
about their dynamics.

The slogan has been
``extract \emph{classical observables} from quantum scattering.''
In particular,
the systematics of this procedure has been 
thoroughly established
by Kosower, Maybee, and O'Connell (KMO'C) \cite{KMOC}.
In the KMO'C formalism,
the starting point is the quantum \textit{S-matrix}.
A formula for classical observables
arises by splitting the S-matrix as
identity plus the $T$-matrix.
One then takes
the $\hbar{\:\to\:}0$ limit
while properly checking the cancellation of divergent terms
dubbed superclassical terms.

Meanwhile, another axis of development has been worldline techniques
\cite{Levi:2018nxp,Porto:2016pyg,Goldberger:2004jt,%
Kalin:2020mvi,Kalin:2020fhe,%
Mogull:2020sak,Jakobsen:2021zvh,Jakobsen:2021smu,Jakobsen:2022psy,Kalin:2022hph,Jakobsen:2023ndj,Jakobsen:2023pvx,%
Dlapa:2021vgp,Dlapa:2022lmu,Dlapa:2023hsl,
Driesse:2024xad,%
Dlapa:2024cje,Dlapa:2025biy,%
Haddad:2025cmw,He:2025how%
}.
In this approach, the astrophysical bodies are modeled directly in terms of point-particle actions.
To obtain classical observables,
one can compute the \emph{on-shell action}
by summing over tree diagrams in the quantum-mechanical perturbation theory.
The on-shell action, also known as principal function,
is a function of the boundary conditions specified for the worldline.
Via the Hamilton-Jacobi framework,
the on-shell action serves as a master object
whose first derivatives
derive the impulse of classical observables.
This inherently describes an \emph{in-out} formalism,
as the definition of the on-shell action
requires a complete specification of the boundary conditions
at the both ends of the worldline.

A more suitable approach for classical physics, however,
could be an \emph{in-in} formalism,
where one deduces the unknown final state 
solely from the knowledge of the initial state.
The systematics of the in-in formalism has been established through works
\cite{Kalin:2020mvi,Kalin:2020fhe,Mogull:2020sak,Jakobsen:2021zvh,Jakobsen:2021smu,Jakobsen:2022psy,Kalin:2022hph,Jakobsen:2023ndj,Jakobsen:2023pvx}.
However, what would be the master object
that derives the classical observables,
in the in-in formalism?
The on-shell action cannot be an answer,
as it is inherently an in-out quantity
due to its very identity as an action.
Namely, \textit{what shall the in-in formalism compute?}

Recently,
works \cite{wlf-ps,ambikerr1,eikonal1,eikonal2,eikonal5,eikonalfield}
have revisited
the foundations of worldline methods
from the perspectives of Hamiltonian (phase space) formulation.
For instance, 
\rcite{wlf-ps} has observed useful universal features of phase space worldline formalisms
while implementing an in-out framework optimized for computing scattering amplitudes.
In the meantime,
\rrcite{eikonal1,eikonal2,eikonalfield,eikonal5}
have investigated the in-in formalism
and identified a clear goalpost:
\emph{scattering generator}.

The earliest inception of this idea
took place in the work \cite{ambikerr1},
where the current author proposed to identify the classical analog of the S-matrix
as a symplectomorphism.
The S-matrix is the map from the initial Hilbert space to the final Hilbert space
in a scattering problem.
In the same way,
the ``\textit{S-symplectomorphism}'' is the map from the initial phase space to the final phase space
in a classical scattering problem.
Arguments were given for a correspondence between the S-matrix and the S-symplectomorphism,
inspired by ideas in symplectic geometry 
and quantization theory
\cite{woodhouse1997geometric,kontsevich}.

An explicit completion of this idea,
however,
was only facilitated
through 
the works \cite{eikonal1,eikonal2}.
There, a precise definition and formula was provided for
the ``logarithm'' of the S-symplectomorphism,
dubbed scattering generator $\chi$.
It is explicitly identified that the scattering generator
is the classical limit of the eikonal matrix \cite{Lehmann:1957zz,Damgaard:2021ipf,Damgaard:2023ttc} $\hat{\chi}$
that arises by 
the exponential parameterization of the S-matrix:
$\hat{S} = \exp(\hat{\chi}/i\hbar)$.
In this sense, the scattering generator $\chi$ is also called the classical eikonal.

Notably, this exponential parameterization facilitates
a computation of the classical observables
via nested Poisson brackets,
in which the superclassical terms are absent by construction.
Hence the slogan may be revised:
``derive classical observables from classical scattering (as S-symplectomorphism).''
This emphasizes that
the classical limit is explicitly imposed from the very beginning.

The nested Poisson bracket formula
provides an alternative to the KMO'C formalism,
systematically deriving the impulse of classical observables 
in a purely in-in fashion
due to the very nature of the exponential map.
As a result,
the scattering generator $\chi$
can serve as the master object in the in-in formalism.
The nested Poisson bracket formula
traces back to \rcite{Damgaard:2023ttc}
and was also independently discovered by 
\rcite{Gonzo:2024zxo}
in systems with symmetries,
just before the release of \rcite{eikonal1}.

In sum,
a well-organized landscape
of worldline techniques
arises,
in terms of the question ``What does it compute?'':
\begin{align*}
	\textbf{In-Out Formalism}
	&\quad\text{computes}\quad
	\textbf{On-Shell Action}
	\,,\\
	\textbf{In-In Formalism}
	&\quad\text{computes}\quad
	\textbf{Scattering Generator}
	\,.
\end{align*}
The perturbative computation of the scattering generator
describes the Magnus \cite{magnus1954exponential} series,
not the Dyson series which the conventional Feynman diagrammatics employs.
Hence
the combinatorics for scattering generator 
differs from that for on-shell action, crucially.

\begin{figure}
	\centering
	\includegraphics[scale=0.95]{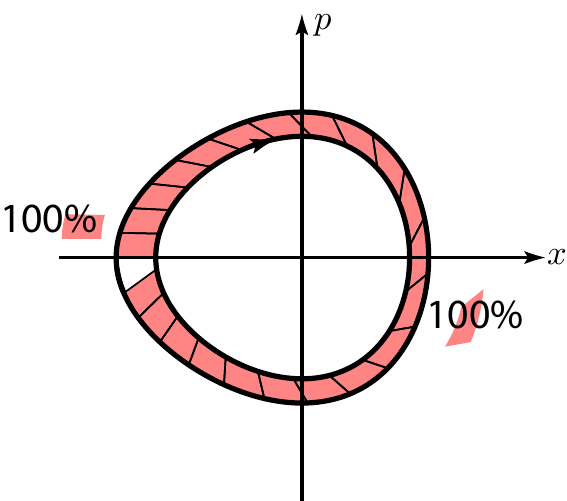}
	\caption{
		An illustration of 
		symplecticity:
		time evolution preserves
		area element in phase space.
	}
	\label{fig:liouville}
\end{figure}

Given the circumstance that our understanding on the scattering generator
has been evolving lately
\cite{ambikerr1,eikonal1,eikonal2,eikonal5,eikonalfield,Haddad:2025cmw},
it might be beneficial if 
the relation between these two currencies for classical observables,
the on-shell action and the scattering generator,
can be clarified.
The purpose of this article is to provide 
a friendly introduction and review
on both currencies,
together with their detailed comparison.
We aim for a purely classical presentation of these ideas
so that
this article might even be used as a class material in a classical mechanics course,
eventually.

The perspective taken in this article 
is the following.
In quantum mechanics and quantum field theory,
unitarity is a sacred principle:
the conservation of probability.
Notably, there exists a classical counterpart of this concept:
\emph{Liouville property}.
Liouville property is the feature of classical time evolution
that it preserves the volume element in phase space,
as is illustrated in \Fig{fig:liouville}.
That is, time evolution is an incompressible flow
in phase space.
This incompressibility directly relates to the classical conservation of probability,
serving as one of the essential keystones in classical statistical mechanics.
From the geometrical perspective
\cite{schutz1980geometrical,arnold1989mathematical,guillemin1990symplectic},
the Liouville property is also referred to as
\emph{symplecticity},
the name of which originates from
the underlying mathematical framework
(symplectic geometry).

If symplecticity is a principle in classical mechanics,
then one would ideally seek for its formulation 
in which symplectivity is manifest.
Amusingly, this angle
precisely brings us to the two protagonists of this article:
the on-shell action and the scattering generator.

The on-shell action ensures symplecticity 
via the Hamilton-Jacobi framework.
The scattering generator, on the other hand,
ensures symplecticity
via the exponential map.
In this article, we will see in detail
how these two approaches differ.
At the practical level,
we will learn through a number of explicit examples
that their values simply do not agree
in general.
At the conceptual level,
we will learn that the on-shell action is much more intricate and delicate.
As an extreme instance,
we will learn at the end of this article that
in some cases the on-shell action is not even defined!
However, the scattering generator is robust
and can be always defined
as long as the equations of motion are consistent.

In this analysis, we will not only consider the scattering context but 
also bulk-to-bulk evolutions,
in which case the logarithm of the time-evolution symplectomorphism is called exponential generator;
this is applicable to bound orbits, for instance.

We will also identify an explicit \emph{matching relation}
between the on-shell action and the exponential/scattering generator:
\eqrefs{matching}{matching.int}.
The crucial idea is that
the latter
can be thought of as an effective Hamiltonian in some sense,
reproducing the entire accumulated effect of the bulk time evolution 
in just ``one second.''
When stated at the level of on-shell actions,
this equivalence stipulates a concrete matching calculation.
We not only provide a proof of this relation
but also explicit checks through multiple examples.

We also try our best to present the idea of the exponential/scattering generator cleanly
and systematically.
Major improvements 
are
emphasis on the strictly (standalone) classical nature of the construction,
exclusive adoption of differential operator and statistical mechanics perspectives,
and elimination of spurious minus signs.

We end with a bullet-point summary of what is achieved in this article:
\begin{itemize}
	\item 
		Systematic derivation of the on-shell action and exponential generator approaches in light of manifest symplecticity.
	\item 
		Explicit demonstrations of how the on-shell action and exponential generator differ.
	\item 
		Explicit matching relation between the on-shell action and exponential generator.
	\item 
		Formulation of classical scattering theory
		strictly within classical mechanics.
\end{itemize}
The essential constructs to be highlighted are
\begin{itemize}
	\item S-symplectomorphism, \eqref{S-Pexp}
		($\leftrightarrow$ S-matrix),
	\item Scattering generator, \eqref{exp.cl.int}
		($\leftrightarrow$ eikonal matrix),
	\item Nested bracket formula, \eqref{expKMOC}
		(alternative to KMO'C).
\end{itemize}

\medskip
\textit{Note added}|%
The relation
between the on-shell action
and the exponential/scattering generator 
is also identified in
\rcite{Kim:2025gis},
released on the same date as this article.

\section{Time Evolution as an Incompressible Flow}
\label{SYMP}

To set the stage,
let us review Hamiltonian mechanics
from the geometrical perspective.

Take a particle in one dimension.
In Newtonian mechanics, 
the particle's dynamics
is described by a second-order differential equation,
say
$m\ddot{x} = F(x)$.
However, an equivalent formulation is viable by a set of first-order differential equations,
say
$\dot{x} = p/m$, $\dot{p} = F(x)$.
This unveils the concept of phase space:
the space $(x,p)$ of position $x$ and momentum $p$.

The idea of phase space
leads to an interesting geometrical reimagination of classical mechanics.
The solution to a set of first-order differential equations
can be interpreted as the flow along a vector field.
Amusingly, 
it turns out that
the time evolution in phase space
is not just an any flow
but a flow that preserves the area element $dp \wedge dx$,
in particular:
see \Fig{fig:liouville} for an illustration.
Thus intuitively,
one can picture
classical time evolution
as the motion of an incompressible fluid streaming through phase space.

This incompressibility is commonly known as the \emph{Liouville property}.
A careful mathematical analysis
shows that it
arises as a vital consistency condition
in classical mechanics.
Especially, 
it describes the very conservation of probability
in classical statistical mechanics.
In this sense, 
the Liouville property describes the classical counterpart of
unitarity in quantum mechanics!

A mathematically precise formulation of these ideas
has long been established in the framework of 
symplectic geometry.
Below, we present a quick summary
without showing every detail of definitions or proofs;
readers may consult to
\rrcite{schutz1980geometrical,arnold1989mathematical,guillemin1990symplectic,woodhouse1997geometric}
for the details.

Symplectic geometry is
the geometry of an invariant area element, $\omega$.
Firstly,
a symplectic manifold $(\P,\omega)$
is an even-dimensional space $\P$
endowed with a symplectic form $\omega$.
A \emph{symplectic form} is a closed, nondegenerate two-form:
\begin{align}
	\label{closure}
	d\omega \,=\, 0
	\,.
\end{align}
Secondly,
let $\Diff(\P)$ be the Lie group of diffeomorphisms in $\P$.
Due to the additional structure $\omega$,
an important subgroup arises in $\Diff(\P)$:
\begin{align}
	\label{SD}
    \Symp(\P,\omega)
        \,\,\subset\,\,
    \Diff(\P)
    \,.
\end{align}
The elements of the Lie group $\Symp(\P,\omega)$
are called \emph{symplectomorphisms}.
These are diffeomorphisms that preserve $\omega$.
Thirdly,
the Lie algebra version of \eqref{SD} reads
\begin{align}
	\label{sd}
    \symp(\P,\omega)
        \,\,\subset\,\,
    \diff(\P)
    \,.
\end{align}
Concretely,
the diffeomorphism Lie algebra $\diff(\P)$ can be 
identified with the space of vector fields in $\P$.
Then elements of $\symp(\P,\omega)$ are vector fields preserving $\omega$,
called \emph{symplectic} \emph{vector fields}.
The Lie bracket in both cases
is the commutator between vector fields as differential operators.

To summarize, a dictionary between mathematics and physics is shown below.
\begin{align}
\begin{split}
    {\renewcommand{\arraystretch}{1.1}
    \begin{array}{rl}
        \text{
            Symplectic Manifold
        }
        &\,\,\,\xleftrightarrow[\,]{}\,\,\,
        \text{
            Phase Space
        }
        \,,\\
        \text{
            Symplectic Form
        }
        &\,\,\,\xleftrightarrow[\,]{}\,\,\,
        \text{
            Invariant Area Element in Phase Space
        }
        \,,\\
        \text{
            Symplectomorphisms
        }
        &\,\,\,\xleftrightarrow[\,]{}\,\,\,
        \text{
            Canonical Transformations
        }
        \,,\\
        \text{
            Symplectic Vector Fields
        }
        &\,\,\,\xleftrightarrow[\,]{}\,\,\,
        \text{
            Infinitesimal Canonical Transformations
        }
        \,.
    \end{array}}
\end{split}
\end{align}

For each function $f$ in phase space,
there is a unique associated symplectic vector field $X_f$.
Geometrically, $X_f$ is a vector field that lies on the contour surfaces of $f$,
which satisfies
\begin{align}
	\label{Xpb}
	X_f[g]
	\,=\,
		\pb{f}{g}
	\,.
\end{align}
Here, $\pb{f}{g}$ is the Poisson bracket,
whose definition follows by
inverting the two-form $\omega$ as a matrix 
at each point in $\P$.
In local patches,
every element of $\symp(\P,\omega)$ arises as $X_f$ for a function $f$.
An important identity reads
\begin{align}
    \label{X-repr}
    [ X_f , X_g ] \,=\, X_\pb{f}{g}
    \,,
\end{align}
establishing a correspondence between
the commutator between vector fields
and the Poisson bracket (Lie algebra homomorphism).
We are omitting the proofs of these facts,
but it should be noted that they
crucially rely on the closure condition in \eqref{closure}.

Lastly,
the symplectic manifold is also equipped with
the Hamiltonian $H$,
a function that holds a special status.
The flow along its symplectic vector field $X_H$
defines the classical time evolution
as a symplectomorphism,
\begin{align}
	\label{U-cl}
    U
        \,\,\in\,\,
    \Symp(\P,\omega)
    \,,
\end{align}
which maps the initial phase space to the final phase space.
This is the mathematical formulation of the earlier statement that
time evolution is an incompressible flow in phase space.
In this paper, we will refer to this $U$ as the
\emph{time-evolution symplectomorphism}.

The bottom line is that
Hamiltonian mechanics
is a machine that
takes the triple $(\P,\omega,H)$
as input
and 
the time-evolution symplectomorphism 
$U \in \Symp(\P,\omega)$
as output.

Finally, it remains to elaborate on
how the Liouville property described earlier
is mathematically formulated.
Suppose the phase space $(\P,\omega)$ has $2n$ dimensions.
The \emph{Liouville measure} is a top-degree differential form defined from the symplectic form as
\begin{align}
	\label{lm}
	\mu
	\,=\,
		\frac{1}{n!}\,
			\underbrace{
				\omega \wedge \cdots \wedge \omega
				{}_{\vphantom{0}}
			}_{\text{$n$ times}}
	\,.
\end{align}
The Liouville theorem
states that
any symplectic vector field $X_f$ in phase space
preserves the Liouville measure in \eqref{lm},
provided the closure condition in \eqref{closure}.
The Liouville property of classical time evolution
is simply the Liouville theorem
applied for the symplectic vector field of the Hamiltonian, $X_H$.
With this understanding, the physical significance of the Liouville measure
is an invariant volume element:
\begin{align}
\begin{split}
    {\renewcommand{\arraystretch}{1.1}
    \begin{array}{rl}
        \text{
            Liouville Measure
        }
        &\,\,\,\xleftrightarrow[\,]{}\,\,\,
        \text{
            Invariant Volume Element in Phase Space
        }
        \,.
    \end{array}}
\end{split}
\end{align}

Crucially,
the Liouville measure
is a vital cornerstone 
in classical statistical mechanics,
facilitating the very counting of states.
On a related note, let $\rho(t)$ be a classical probability distribution (a normalizable smooth function in $\P$)
that is transported by the classical time evolution.
As is well-known,
$\rho(t)$ is governed by
the \emph{Liouville equation}:
\begin{align}
	\label{Leq}
	\dot{\rho}(t)
	\,=\,
		X_{H(t)}[\rho(t)]
	\,.
\end{align}
Here, the overdot denotes $\partial/\partial t$,
and a time-dependent Hamiltonian is assumed for full generality.
\eqref{Leq} is a partial differential equation,
where $X_{H(t)}$ is understood as a first-order differential operator.
A classical observable $\O$ is a smooth function in phase space.
Given a classical probability distribution $\rho(t)$,
the expectation value of $\O$ is defined as
\begin{align}
	\label{def-expval}
		\int_\P\,\hem \O\mem \rho(t)\mem \mu
	\,,
\end{align}
which describes the integration of a top-degree differential form
$\O\mem \rho(t)\mem \mu$
over the entire phase space $\P$.
The invariance of $\mu$ plays crucial roles in this framework,
ensuring consistency of the statistical/probabilistic interpretation.
For instance, it holds that 
$\rho(t)$ is normalized at all times
if it was normalized once at a particular time $t = t_0$,
by the very invariance of $\mu$ under time evolution.

This concludes our crash course on the geometrical formulation of Hamiltonian mechanics,
which yet showed only the strictly necessary equations.
For a further clarification, we remark that
the three statements below are mathematically equivalent
\cite{schutz1980geometrical,arnold1989mathematical,guillemin1990symplectic,woodhouse1997geometric}.
\begin{enumerate}
\item 
	Closure of the area element, $d\omega = 0$.
\item 
	Time evolution preserves the area element $\omega$.
\item 
	Time evolution preserves the volume element $\mu$.
\item 
	Local existence of a one-form $\theta$ such that $\omega = d\theta$.
\end{enumerate}
The first two statements are sometimes referred to as \emph{symplecticity} of classical mechanics.
Reiterating,
this symplecticity
is equivalent to the Liouville property above:
\begin{align}
	\text{Symplecticity}
	\,\,\,=\,\,\
	\text{Liouville Property}
	\,\,\,=\,\,\,
	\text{Classical Probability Conservation}
	\,.
\end{align}

Meanwhile,
the fourth statement introduces the one-form $\theta$,
called \emph{symplectic potential}.
Given the Hamiltonian $H$ and a symplectic potential $\theta$,
one can immediately write down a first-order action.
In this sense, symplecticity
means that the physical system is Lagrangian.
It is not difficult to see that
the converse is also true.

This discussion should justify
taking symplecticity
as a principle in classical mechanics.

\section{Currency for Manifest Symplecticity}
\label{CURR}

The bottom line of \Sec{SYMP}
is that
Hamiltonian mechanics
in a phase space $(\P,\omega)$
is a machine that outputs 
the time-evolution symplectomorphism, $U \in \Symp(\P,\omega)$.
Physicists, as deeply practical beings, 
desire to compute this symplectomorphism \textit{explicitly}.
However, what does it mean to ``compute'' a symplectomorphism or even a diffeomorphism?

Geometrically,
there are two ways to represent\footnote{
	Note that
	one can take the wording ``represent'' here
	as the precise technical term exploited in the mathematical theory of representations
	\cite{vershikgelfand1975diffreps},
	if wanted.
} a diffeomorphism.
The first is to directly describe the trajectory of each \emph{point} in the manifold.
This is indeed the point of view presumed in \Sec{SYMP}.
The second, however,
is to state how each \emph{test function} in the manifold
is transported by the flow.

Recall our earlier fluid analogy in \Sec{SYMP}.
The first approach observes how a tiny bead of neutral buoyancy travels with the fluid.
The second approach, in contrast,
sprinkles a non-dissipating ink in the fluid
and observes how its shape evolves.
These two descriptions are equivalent.
In particular,
suppose a test function sharply peaked around a center.
Then one cannot tell if one is looking at a point or a tiny splash of ink.

The astute reader will point out that the second approach
is natural in the context of classical statistical mechanics,
in which case the time-evolution symplectomorphism $U$
is represented as
the \emph{differential operator}
that maps the initial classical probability distribution
to the final classical probability distribution.
In this paper, we will adopt this perspective.
For instance, \Sec{SOLVING} will derive
explicit formulae computing
$U$ as a differential operator,
by studying the Liouville equation in \eqref{Leq}.
Note that $\rho(t)$ is precisely the mathematical model
for the time-evolving profile of the non-dissipating ink.

With these preliminary remarks,
we invite the reader to a hypothetical agora
where an animated debate on the following question is in full swing.
\begin{center}
	If symplecticity is a principle in classical physics,\\
	in what data type shall the output, $U$,
	be presented?
\end{center}
The context is that
one would ideally aim for a formulation of classical mechanics
in which symplecticity is manifest,
given its physical significance highlighted in \Sec{SYMP}.
For simplicity,
the scope is restricted to phase-space-based formulations.

Below, 
we are introduced to
three candidates
who are thrilled to answer this question
from their own respective perspectives.

The first candidate 
advocates a straightforward approach.
Pick a point in phase space
and let it flow under the classical time evolution.
State its final coordinates
as a function of the initial coordinates.
This specifies the map $U$.

However, this approach is too naive,
neither manifesting nor emphasizing symplecticity as the very crucial feature of the phase space framework.
Namely, this approach presents $U$ simply as a \emph{diffeomorphism element}, $U \in \Diff(\P)$.
This does not establish that $U$ actually belongs to the smaller group, $\Symp(\P)$.

Now the second candidate enters the screen
and argues that
historically, 
more sophisticated formulations existed and were motivated
\textit{because} the first candidate's method did not work.
The second candidate invokes 
what is known as
the Hamilton-Jacobi formalism,
commonly covered in textbooks \cite{goldstein2002classical,arnold1989mathematical}.
Mathematically, this framework
encodes the relation between the initial and final variables
in the so-called principal function $F$
defined on a product manifold $\L' \mtimes \L$,
such that symplecticity is guaranteed.
Here, $\L$ and $\L'$
are Lagrangian submanifolds
in the initial and final phase spaces, respectively.
Lagrangian submanifolds are
submanifolds of maximal dimension in a symplectic manifold
on which the Poisson brackets vanish:
submanifolds of fixed positions or momenta, for instance.
This sounds quite complicated though,
but physically,
the principal function $F$
is simply the \emph{on-shell action}
for a chosen boundary condition.
Thus instead of invoking the sophisticated terminology ``Lagrange submanifolds,''
one could have just said that
the computation of the on-shell action
fixes either positions or momenta
at each end 
by its very nature.

This might sound like a reasonable response.
However,
despite being widely acknowledged,
the Hamilton-Jacobi formalism
still entails some critical disadvantages.
In classical mechanics,
it is virtually always the case that
one is given with the knowledge of only the initial conditions.
Unfortunately,
the principal functions
intrinsically are
functions of
both initial and final variables,
which should be clear from their very identity as on-shell \textit{actions}.
Hence 
ironically,
one has to invoke some of the final scattering 
variables
before even solving
the very time evolution problem
whose goal is to determine the unknown final
data.
To recapitulate,
the practical situations in classical mechanics
are always
\textit{in-in} boundary value problems,
whereas 
the Hamilton-Jacobi framework
is inherently an \textit{in-out} formalism.

Finally, 
the scene cuts to a close-up shot of the third candidate,
who boldly proposes a new approach.
The idea is yet remarkably simple:
write
the time-evolution symplectomorphism $U$
as the \emph{exponentiation of a symplectic vector field}.
That is,
\begin{align}
    \label{exp.cl}
    U
        \mem=\,
    \exp(X_G)
    \,.
\end{align}
It is evident that this explicitly guarantees the symplecticity of $U$:
\begin{align}
	\label{expmapG}
	X_G \in \symp(\P,\omega)
	\qiq
	U \in \Symp(\P,\omega)
	\,.
\end{align}
\eqref{expmapG}
describes the exponentiation of a Lie algebra element
to a Lie group element.

Physically, this proposal
well-adapts to the in-in boundary value problems typical in classical physics.
By the very nature of the exponential map,
\eqref{exp.cl} implements the action of $U$
in terms of a Taylor series acted on the initial phase space.
Hence
symplecticity
is manifested
in an in-in formulation.
Mathematically, this proposal is also significantly simpler than the Hamilton-Jacobi treatment.
The notion of Lagrangian submanifolds
is completely unnecessary.
All coordinates in the phase space
are strictly put on an equal footing
and
no discrimination exists at all.

In the final scene,
the judges 
proclaim the third candidate as the victor
and identify \eqref{exp.cl}
as the sought-after formulation of classical mechanics
in which symplecticity stands manifest.
Crucially,
the function $G$ in \eqref{exp.cl}
is the generator of time evolution
in the Poisson and exponential map sense.
Namely, an \emph{exponential generator}.
The judges promote $G$ to a new currency
for manifest symplecticity in classical mechanics,
overturning the traditional, long-standing monopoly
of the on-shell action $F$.

The above script is just a hypothetical scenario after all,
intended 
for pedagogical purposes.
However, we would still like to appreciate the lesson that
the simplest method of manifesting symplecticity
is to employ the exponential map representation in \eqref{exp.cl}
in the in-in setup.
Accordingly,
in this paper, we spotlight the exponential generator $G$
as a novel idea that deserves attention.
It is rather surprising that
the precise concept of $G$ emerged only recently
through works \cite{ambikerr1,eikonal1,eikonal2}
as mentioned in the introduction.

\subsection{Example: Falling Apple}
\label{CURREX}

To clarify the points made above to a further extent,
let us investigate an explicit example:
a falling apple.
Despite its simplicity,
this innocuous example
teaches us a valuable lesson that
the on-shell action $F$ and the generator $G$ are completely different objects,
not only conceptually:
plugging in the explicit solutions
even fails to bring them to the same value.

The falling apple is a Hamiltonian system $(\mathbb{R}^2, \omega, H)$,
where
\begin{align}
	\label{apple}
	\omega \,=\,
		dp \wedge dx
	\,,\quad
	H(x,p) \,=\,
		\frac{p^2}{2m} + mgx
	\,.
\end{align}
The constant parameter $g$ describes
the uniform gravitational field.
The canonical Poisson bracket reads $\pb{x}{p} = 1$.
The Hamiltonian equations of motion are
\begin{align}
\begin{split}
    \label{eom-apple}
    \dot{x}(t) \,=\, p(t)/m
    \,,\quad
    \dot{p}(t) \,=\, - mg
    \,.
\end{split}
\end{align}
Let $U$ be the time-evolution symplectomorphism
for this apple,
from time $t = 0$ to $t = T$,
say.
Our goal is to compute $U$ in various data types
and make a comparison between them.

\subsubsection{Diffeomorphism}
\label{CURREX.diff}

We start with the first candidate's proposal:
obtain $U$ as a diffeomorphism.
Surely, this is the most straightforward and rudimentary approach.
By solving the equations of motion in \eqref{eom-apple}, we find
\begin{align}
    \label{apple-U}
    (x,p)
    \,\,\,\mapsto\,\,\hem
    (x',p')
    \,=\,
    \bigg(\,{
        x 
        + \frac{p}{m}\, T
        - \frac{1}{2}\, g\hem T^2
        \,,\,
        p - mg\hem T
    }\mem\bigg)
    \,.
\end{align}
Unfortunately,
when stated in this format,
symplecticity
is only 
assured
after
an additional 
calculation:
\begin{align}
    dp' \wedge dx'
    \,=\, 
    dp \mem\wedge \nem\bigg(\,{
        dx + \frac{T}{m}\mem dp
    }\mem\bigg)
    \,=\,
    dp \wedge dx
    \,.
\end{align}

As a differential operator, $U$ is given by
\begin{align}
\begin{split}
	\label{apple-U-diffop}
	U
	\,=\,
		1 
		&+ T\,\bb{\nem
			- \frac{p}{m}\, \frac{\partial}{\partial x}
			+ mg\, \frac{\partial}{\partial p}
		}
	\\
		&+ \frac{T^2}{2}\,\bb{
			\frac{p^2}{m^2}\, \frac{\partial^2}{\partial x^2}
			+ m^2g^2\, \frac{\partial^2}{\partial p^2}
			- 2gp\, \frac{\partial^2}{\partial x\hem \partial p}
			- g\, \frac{\partial}{\partial x}
		}
		+ \O(T^3)
	\,,
\end{split}
\end{align}
which can be obtained by 
transporting a test function by \eqref{apple-U}
and then Taylor expanding.
Unfortunately,
it is hard to tell directly from \eqref{apple-U-diffop}
if $U$ really describes a symplectomorphism
or even a diffeomorphism.
This is because not all differential operators
originate from a diffeomorphism.
That is,
writing down any differential operator
would not guarantee that you have found
a Taylor expansion for shifting coordinates:
a ``bitter life lesson,'' to say.
Surely, 
we will see shortly that the exponential parameterization in \eqref{exp.cl}
nicely circumvents this issue.

\subsubsection{On-Shell Action}
\label{CURREX.F}

Next, we move on to the second candidate's proposal:
Hamilton-Jacobi formalism.
To this end, we may want to briefly review the fundamentals of the formalism
before diving into explicit computations.

From a mathematical standpoint,
the idea of the Hamilton-Jacobi formalism is to ensure symplecticity
by finding
the difference between two symplectic potentials
for the initial and final phase spaces
as an (locally) exact one-form $dF$,
where $F$ is the principal function.
As briefly mentioned in \Sec{SYMP},
a symplectic potential
is a one-form whose exterior derivative gives the symplectic form.
For our falling apple,
we have $\omega = dp \wedge dx$,
so
the natural choices for the symplectic potential are
$p\mem dx$ and $-x\mem dp$.
From these options,
the following four versions of the principal function arise:
\begin{align}
\begin{split}
	\label{dF}
    dF^{(1)} = p' dx' - p\mem dx
    &\,,\quad
    dF^{(2)} = - x' dp' - p\mem dx
    \,\\
    dF^{(3)} = p' dx' + x\mem dp
    &\,,\quad
    dF^{(4)} = - x' dp' + x\mem dp
    \,.
\end{split}
\end{align}
These four realizations
are all related to each other
via Legendre transformations.
It follows that they provide maps
between products of Lagrangian submanifolds:
\begin{align}
\begin{split}
    \label{mapF}
    F^{(1)}
        \,\,:\,\,
    (x',x)
        \,\,\mapsto\,\,
    (p',p)
    &\,,\quad\,
    F^{(2)}
        \,\,:\,\,
    (p',x)
        \,\,\mapsto\,\,
    (x',p)
    \,,\\
    F^{(3)}
        \,\,:\,\,
    (x',p)
        \,\,\mapsto\,\,
    (p',x)
    &\,,\quad\,
    F^{(4)}
        \,\,:\,\,
    (p',p)
        \,\,\mapsto\,\,
    (x',x)
    \,.
\end{split}
\end{align}
The crucial point is that
acting on an exterior derivative to \eqref{dF} gives
\begin{align}
	\label{HJsymp}
	\text{\eqref{dF}}
	\qiq
		dp' \wedge dx' \,=\, dp \wedge dx
	\,,
\end{align}
so the preservation of the symplectic form is manifest.

From the physical standpoint,
however,
the principal functions
are
nothing but on-shell actions
with proper boundary terms
as is remarked earlier.
For example,
the second principal function in \eqref{dF} is given by
the on-shell (extremum) value
of a phase space action
with a boundary term $-p(T)\mem x(T)$:
\begin{align}
	\label{F2}
	F^{(2)}(p',x)
	\,=\,
	\underset{\substack{
		p(T) = p'\\
		x(0) = x
	}}{\Ext}\,
	\bbsq{
	    -p(T)\mem x(T)
	    +\hnem
	    \int_{0}^{T} dt\,\,
	    \BB{
	    	p(t)\mem \dot{x}(t) {\,-\,} H(x(t),p(t);t)
	    }
	}
    \,.
\end{align}
Here,
the computation of the right-hand side 
must fix
the initial position $x$ and the final momentum $p'$,
which is the very boundary condition stipulated by the variational principle.
With this point understood,
it is easily checked that 
the on-shell variation of \eqref{F2}
reads
$\delta F^{(2)}(p',x)
= -x(T)\mem \delta p(T) - p(0)\mem \delta x(0)$,
deriving the differential relation stipulated in \eqref{dF}.
If a different boundary condition is employed,
the boundary action changes accordingly
so that the Legendre transformations
are precisely reproduced.
In this perspective, the idea of Lagrangian submanifolds
arises naturally
from the boundary conditions for a phase space action.

Now we can switch to explicit evaluations.
The second principal function in \eqref{mapF}
for our falling apple
is explicitly given as
\begin{align}
	\label{F2-apple}
    F^{(2)}(p',x)
    \,=\,
        - p'\hem x 
        - \frac{{p'}^2}{2m}\, T
        - mg\, \bigg(\mem{
            x\mem T
            + \frac{p'}{2m}\, T^2
        }\mem\bigg)
        - \frac{mg^2}{6}\mem T^3
    \,.
\end{align}
By referring to the differential relation $dF^{(2)} = -x'\mem dp' - p\mem dx$ in \eqref{dF},
one then finds
$x' = -\partial F^{(2)}\nem/\partial p'$ and $p = -\partial F^{(2)}\nem/\partial x$
as functions of $p'$ and $x$.
In this sense, 
$F^{(2)}$ provides a map $(p',x) \mapsto (x',p)$
as stated in \eqref{mapF}.
Explicitly,
\begin{align}
\begin{split}
	\label{HJ1D}
	x'
	\,&=\,
		-\frac{\partial F^{(2)}}{\partial p'}
	\,=\,
		x + \frac{p'}{2m}\mem T
		+ \frac{1}{2}\, g\hem T^2
	\,,\\
	p
	\,&=\,
		-\frac{\partial F^{(2)}}{\partial x}
	\,=\,
		p' + mg\hem T
	\,.
\end{split}
\end{align}
This approach guarantees symplecticity:
$0 = ddF^{(2)} = d( -x' dp' - p\mem dx) = dp' \wedge dx' - dp \wedge dx$,
as is shown in \eqref{HJsymp}.
Unfortunately, however,
the map $(x,p) \mapsto (x',p')$
in \eqref{apple-U}
can be found
only after 
the process of
re-expressing $x'$ and $p'$ as functions of $x$ and $p$,
which can be involved in general.

\subsubsection{Exponential Map}
\label{CURREX.G}

Finally, we consider
the exponential map in \eqref{exp.cl}.
As the Hamiltonian $H$ in \eqref{apple} is time-independent,
the generator $G$ is simply the $H$ times the total elapsed time:
\begin{align}
	\label{G-apple}
    G(x,p)
    \,=\,
        \bigg(\,{
            \frac{{p}^2}{2m} + mgx
        }\bigg)\mem 
        T
    \,.
\end{align}
To reproduce \eqref{apple-U},
one computes \textit{nested} Poisson brackets as
\begin{align}
	\label{getGexp}
{\renewcommand{\arraystretch}{0.5}
\begin{array}{lll}
    x'
    &\,=\,
    \displaystyle
        x + \pb{x}{G} + \frac{1}{2!}\mem \pb{\pb{x}{G}}{G} + \cdots
    &\,=\,
    \displaystyle
        x + \frac{p}{m}\mem T
        - \frac{1}{2}\, g\hem T^2
    \,,\\
    \\
    p'
    &\,=\,
    \displaystyle
        p + \pb{p}{G} + \frac{1}{2!}\mem \pb{\pb{p}{G}}{G} + \cdots
    &\,=\,
        p - mg T
    \,.
\end{array}}
\end{align}
To reproduce \eqref{apple-U-diffop},
one exponentiates $X_G$
as a differential operator:
\begin{align}
	\label{XG-apple}
	X_G
	\,=\,
		T\,\bb{\nem
			- \frac{p}{m}\, \frac{\partial}{\partial x}
			+ mg\, \frac{\partial}{\partial p}
		}
	\qiq
	U \,=\, \exp(X_G)
	\,.
\end{align}
In this approach,
symplectivity of $U$ is manifest
because \eqref{XG-apple} is the very symplectic vector field of \eqref{G-apple}.
Moreover, the final variables are deduced from the knowledge of only the initial variables:
an in-in formalism.

\subsection{Exponential Generator $\neq$ On-Shell Action}
\label{GvsF}

An important remark is
that $G$ and $F$ are different,
not only conceptually but also in terms of their explicit values.
Naively, one might conceive a thought that $G$
and $F$
are similar:
they have the same dimensions (units of angular momentum, i.e., action)
and are both things that would make an appearance in some advanced course in classical mechanics.
In fact,
one might even confuse between them
and {misleadingly call $G$ as on-shell action}!

However, even the simple example of an innocuous falling apple
reveals their, grossly, diverging identities.

\subsubsection{Practical Differences}

Take the exponential generator $G$ in \eqref{G-apple}.
Take the on-shell action $F^{(2)}$ in \eqref{F2-apple}.
Let us compare their explicit values
by plugging in the relations between $x,p,x',p'$ found in \eqref{apple-U}.
By plugging in $p' = p-mgT$ in \eqref{F2-apple}, we find
\begin{align}
	\label{F2-apple.eval}
	F^{(2)}
	\,=\,
		-p\mem x - \frac{p^2}{2m}\mem T
		+ \frac{1}{2}\, gp\mem T^2
		- \frac{1}{6}\, mg^2\mem T^3
	\,.
\end{align}
Evidently, this is not equal to $G$ in \eqref{G-apple}
(nor $-G$).
While \eqref{G-apple} is at most linear in both $g$ and $T$,
\eqref{F2-apple.eval} is nonlinear in both $g$ and $T$.

With this failure, we try a different boundary condition.
For instance, the on-shell action fixing positions 
is defined as the extremum value
\begin{align}
	\label{F1}
	F^{(1)}(x',x)
	\,=\,
	\underset{\substack{
		x(T) = x'\\
		x(0) = x
	}}{\Ext}\,
	\bbsq{\,\hnem
	    \int_{0}^{T} dt\,\,
	    \BB{
	    	p(t)\mem \dot{x}(t) {\,-\,} H(x(t),p(t);t)
	    }
	}
    \,.
\end{align}
For our falling apple, it is found as
\begin{align}
	\label{F1-apple}
	F^{(1)}(x',x)
	\,&=\,
		\frac{m}{2T}\,
			(x'\nem-x)^2
		- mg\mem T\,
			\frac{x'\nem+x}{2}
		- \frac{1}{24}\,
			mg^2\mem T^3
	\,,
\end{align}
which, on plugging in $x' = x + (p/m)\mem T - gT^2/2$ from \eqref{apple-U},
becomes
\begin{align}
	F^{(1)}
	\,=\,
		\frac{p^2}{2m}\, T
		- mgT\mem x
		- gp\mem T^2
		+ \frac{1}{3}\, mg^2\mem T^3
	\,.
\end{align}
Evidently, this is also not equal to $G$ in \eqref{G-apple}.
In fact, any of the other instances of the principal function
fails to be equated with $G$
in terms of explicit values.

Therefore, if one computed $G$ for a physical system,
one did not compute any instance of an on-shell action.

\subsubsection{Conceptual Differences}

The explicit mismatch above is already clear,
but there are further differences between $F$ and $G$ at the conceptual levels as well.

\begin{enumerate}
\item 
	$F$ is inherently an in-out quantity since it is an action.
	$G$ is inherently in-in, however.
\item
	Legendre transformations can never bring $F$ to a $G$.
	Legendre transformations can never put canonically conjugate variables together.
	Physically speaking,
	$x$ and $p$ cannot be simultaneously put together on the same side
	as a boundary condition for a phase space action.
	(Quantum mechanically, this is due to the very uncertainty principle.)
	$G$ is not related to
	any of the on-shell actions
	via Legendre transformations.
\item 
	Consider how each formalism deduces the unknown variables.
	With $F$, one computes just a \textit{single} derivative as in \eqref{HJ1D}.
	With $G$, one adds up \textit{nested} Poisson brackets as in \eqref{getGexp}, however.
\item 	
	$F$ and $G$ are also different in terms of the geometric structures they require.
	$F$ requires a symplectic manifold with polarization structures (Lagrangian submanifolds).
	$G$ requires only a Poisson manifold.
	
	Physically speaking, $F$ requires an action principle because it is literally an action.
	$G$, on the other hand, is defined straight out of classical equations of motion
	and never has demanded an action principle.
	Hence there is no boundary condition (Lagrange submanifolds) to talk about, after all.
	One simply needs a Poisson bracket.
	
	Later in \Sec{MAGSPIN}, this point will be elucidated by a concrete physical example.
\item 
	$F$ is an action, so its perturbative computation admits a path integral approach
	(summing over tree Feynman graphs).
	However, 
	it appears that there might be no good path integral representation for the perturbative computation of $G$.
	
	The systematics for the former 
	has been worked out in \rcite{wlf-ps}:
	phase space Feynman rules in the in-out setup.
	The combinatorics is that of the usual Feynman graphs.
	In contrast,
	the diagrammatics for $G$ has been worked out in \rcite{eikonal2}.
	The combinatorics is quite different, involving even a use of Hopf-algebraic methods.
\end{enumerate}

\subsection{Matching}
\label{MATCHING}

Still, however,
one could hope identifying
a concrete relation between the $G$ and $F$.
Regarding this curiosity,
we provide another perspective toward $G$:
$G$ may be thought of as
an effective Hamiltonian.
\eqref{exp.cl} states that
the entire time evolution 
from the initial time
to the final time
by the Hamiltonian $H(t)$,
which is generally time-dependent,
is directly reproduced by a steady evolution by $G$
for a unit time.
In this sense, $G$ is an effective Hamiltonian
encapsulating the entire history
within a unit-time flow.

Based on this effective equivalence 
at the level of
time evolution,
one could argue that the on-shell actions arising from the two descriptions
should also match:
\begin{align}
\begin{split}
	\label{matching}
	&
	\underset{\substack{
		x(\tf) = x'\\
		x(\ti) = x
	}}{\Ext}\,
	\bbsq{
    \int_{\ti}^{\tf} dt\,\,
    	\BB{
	    	p(t)\mem \dot{x}(t) {\,-\,} H(x(t),p(t);t)
    	}
    }
    \\
   	&
	=\,
	\underset{\substack{
		x(1) = x'\\
		x(0) = x
	}}{\Ext}\,
	\bbsq{
    \int_{0}^{1} dt\,\,
    	\BB{
	    	p(t)\mem \dot{x}(t) {\,-\,} G(x(t),p(t))
    	}
	}
    \,.
\end{split}
\end{align}
Here, we have simply assumed the boundary condition fixing positions at both ends
($F^{(1)}$ in \eqref{dF}),
as other cases will be implied through Legendre transformations.
We have also assumed generic initial and final times, $\ti$ and $\tf$.

To clarify,
the left-hand side of \eqref{matching} 
computes the on-shell action from the original description given by the Hamiltonian $H(t)$.
The right-hand side of \eqref{matching},
on the other hand,
computes the on-shell action in the effective description due to $G$.
The statement is that they
match
as principal functions
implementing the same effect on the boundary phase spaces.
As is tested in \Sec{EXAMPLES},
this assertion holds true.
For time-independent systems such as the falling apple,
\eqref{matching} is trivial by a reparameterization.
See \rcite{Kim:2025gis} as well.

In fact, 
a proof of \eqref{matching} 
could appeal to
quantum mechanics.
Firstly, \eqref{matching} is lifted to the fully quantum level 
by replacing the extremum operators to path integrals $\int \mathcal{D}x\mem \mathcal{D}p$.
Then the left-hand side of \eqref{matching} computes a transition amplitude
$\langle x' |\mem \hat{U} \hem| x \rangle$,
where $\hat{U}$ is the time-evolution unitary operator
from $t {\:=\:} \ti$ to $t {\:=\:} \tf$.
A quantum-mechanical version of \eqref{exp.cl}
reads $\hat{U} = \exp(\hat{G}/i\hbar)$.
Hence this transition amplitude
boils down to $\langle x' |\hem \exp(\hat{G}/i\hbar) \mem| x \rangle$,
yielding the right-hand side of \eqref{matching}.
Eventually, we take the classical limit 
and reproduce \eqref{matching}.

\section{Computation of Time-Evolution Symplectomorphism}
\label{SOLVING}

Having established the fundamentals,
we now derive explicit formulae computing
the time-evolution symplectomorphism
as a differential operator,
by studying the Liouville equation in \eqref{Leq}.
For simplicity,
we set the initial and final times as $t = 0$ and $t = T$,
which is a totally arbitrary choice,
to clarify.

\subsection{Dyson Series}

Given the understanding that
the vector field $X_{H(t)}$ in \eqref{Leq} 
is taken as a differential operator,
the solution
to the Liouville equation in \eqref{Leq}
is immediate by the well-known 
\emph{Dyson series} \cite{Dyson:1949ha}
formula:
\begin{align}
\begin{split}
	\label{dyson-rho}
	\rho(T)
	\,&=\,
		\Pexp{\,
			\int_{0}^{T} dt\,\,
				X_{H(t)}
		}\,
		\rho(0)
	\,.
\end{split}
\end{align}
From \eqref{dyson-rho},
we directly read off the time-evolution symplectomorphism as a differential operator,
\begin{align}
	\label{U-Pexp}
	U
	\mem=\,
		\Pexp{\,
			\int_{0}^{T} dt\,\,
				X_{H(t)}
		}
	\,\,\,:\,\,\,
		\rho(0)
	\,\,\mapsto\,\,\hnem
		\rho(T)
	\,.
\end{align}
The ordered exponential unpacks into the infinite sum
\begin{align}
\begin{split}
	\label{Usol}
	U
	\mem&=\,
		1 
		\,+\,\mem
	    {
	        \sum_{n=1}^\infty\,
				\int_{t_1 > t_2 > \cdots > t_n}\nem d^nt
	                \,\, X_{H(t_1)} X_{H(t_2)} \cdots X_{H(t_n)}
	    }
	\,,\\
	\mem&=\,
		1
		+
		\int_0^T dt_1\,\,
			X_{H(t_1)}
		+
		\int_0^T dt_1
		\int_0^{t_1} dt_2\,\,
			X_{H(t_1)}
			X_{H(t_2)}
		+
		\cdots
	\,,
\end{split}
\end{align}
where each integration variable $t_k$ runs through $[0,T]$
while $d^nt$ abbreviates $dt_1\hem dt_2\hem \cdots\hem dt_n$.

\newpage

In general, \eqref{Usol}
evaluates to a giant sum of differential operators of all higher orders.
For instance,
$X_{H(t_1)} X_{H(t_2)} \cdots X_{H(t_n)}$ in \eqref{Usol}
is in general
an $n$\textsuperscript{th} order differential operator
$\rho \mapsto X_{H(t_1)}[ X_{H(t_2)}[ \cdots X_{H(t_n)}[\rho] ] ]$.
For $n {\,>\,} 1$, this is not algebra-valued in general,
meaning that it does not describe an element of 
$\symp(\M,\omega)$
(nor $\diff(\M,\omega)$).

As a result, the Dyson series formula in \eqref{Usol}
grossly obscures the fact that
$U$ originated from a diffeomorphism/symplectomorphism in phase space.
It rather returns a giant differential operator,
which seems so generic that
it is left unclear
why it has to describe the exponentiation of a vector field:
the bitter life lesson.

\subsection{Magnus Series}
\label{MAGNUS}

To this end,
consider the
exponential formula in \eqref{exp.cl}:
$U = \exp(X_G)$.
When applied to \eqref{U-Pexp},
we find that 
the symplectic vector field
$X_G$ computes a log of an ordered exponential:
\begin{align}
	X_G
	\,=\,
		\log\mem
			\Pexp{\,
				\int_{0}^{T} dt\,\,
					X_{H(t)}
			}
	\,.
\end{align}
Luckily, it has been well-established that
the log of an ordered exponential
admits a unique series expansion,
known as the \emph{Magnus series} \cite{magnus1954exponential}.

At low orders, the Magnus series reads
\begin{align}
	\label{XGsol}
	X_G
	\,=\,
		&
		\int_0^T dt_1\,\,
			X_{H(t_1)}
		+
		\frac{1}{2}\,
		\int_0^T dt_1
		\int_0^{t_1} dt_2\,\,
			[ X_{H(t_1)} , X_{H(t_2)} ]
		+
		\cdots
	\,.
\end{align}
Compare this with the second line in \eqref{Usol}.
The crucial property of Magnus series is that
each term is 
constructed in terms of the Lie bracket
and thus is manifestly algebra-valued.
That is, \eqref{XGsol} computes a symplectic vector field $X_G \in \symp(\P,\omega)$
by a series expansion
where each term is a symplectic vector field by itself.

Using the very Lie algebra homomorphism in \eqref{X-repr},
\eqref{XGsol} translates to
\begin{align}
	\label{XGsol-pushed}
	X_G
	\,=\,
		&
		\int_0^T dt\,\,
			X_{H(t)}
		+
		\frac{1}{2}\,
		\int_0^T dt_1
		\int_0^{t_1} dt_2\,\,
			X_{\pb{H(t_1)}{H(t_2)}}
		+
		\cdots
	\,.
\end{align}
This implies that
the generator $G$ is uniquely determined up to additive constants as
\begin{align}
	\label{Gsol}
	G
	\,=\,
		&
		\int_0^T dt\,\,
			H(t)
		+
		\frac{1}{2}\,
		\int_0^T dt_1
		\int_0^{t_1} dt_2\,\,
			\pb{H(t_1)}{H(t_2)}
		+
		\cdots
	\,,
\end{align}
where we have used the property that
$X_f + X_g = X_{f+g}$.
The straightforward correspondence between \eqref{XGsol}
and \eqref{Gsol} should be clear.
Note that $f \mapsto X_f$ projects out additive constants in $f$.

The charm of this approach is that
it suffices to handle scalar functions,
which should be clear from the final formula in \eqref{Gsol}.
Computing the Dyson series in \eqref{Usol},
in contrast,
requires handling differential operators of all higher orders,
which is usually not a practical option.

To find higher-order terms in the Magnus series,
one can use
a nice recursion relation based on Bernoulli numbers,
provided by Magnus \cite{magnus1954exponential} himself.
The $k$\textsuperscript{th} Bernoulli number, $B_k$,
is defined as
\begin{align}
	\label{bernoulli}
	\frac{\xi}{\mathe^\xi - 1}
	\,=\,
	\sum_{k=0}^\infty\,
	\frac{B_k}{k!}\,
	    \xi^k
	\qiq
	B_0 = {1}
	\,,\,\,\,
	B_1 = -\frac{1}{2}
	\,,\,\,\,
	B_2 = {\frac{1}{6}}
	\,,\,\,\,
	B_3 = 0
	\,,\,\,\,
	\cdots
	\,.
\end{align}
The generator $G$
then follows as the sum
$G = \sum_{n=1}^\infty \G_n(T)$,
where
\begin{subequations}
\label{recursion.magnus}
\begin{align}
    \label{recursion.magnus.1}
    \dot{\G}_1(t) 
    &\,=\, H(t)
    \,,\\
    \label{recursion.magnus.n}
    \dot{\G}_n(t)
    &\,=\,
    \sum_{k=1}^\infty\,
        \frac{B_k}{k!}
        \kern-0.1em
        \sum_{\a_1+\cdots+\a_k = n-1}\nem\hnem
   			\pb{\G_{\a_1}\nem(t)}{
   				\pb{\G_{\a_2}\nem(t)}{
   					\cdots
   						\pb{\G_{\a_k}\nem(t)}{
   							H(t)
   						}
   				}
   			}
   	\quad
   	(n>1)
	\,.
\end{align}
\end{subequations}
Here, $\a_1,\cdots,\a_k$ are positive integers.
The initial condition for integrating \eqref{recursion.magnus} is $\G_n(0) = 0$ for all $n$.
The details are left to \App{MAGPF},
together with an alternative formula \cite{strichartz1987campbell}
that could be also useful.

The above recursive method
concretely and uniquely defines the Magnus series
to all orders.
For reference,
below we explicitly spell out the Magnus series for the generator $G$
up to the fourth order:
\begin{align}
\begin{split}
	\label{G+4}
	G
	\,=\,
		&
		\int dt_1\,\,
			H(t_1)
		+
		\frac{1}{2}\mem
		\int_{t_1>t_2}\nem d^2t\,\,
			\pb{H(t_1)}{H(t_2)}
\\
		&
		+
		\frac{1}{6}\mem
		\int_{t_1>t_2>t_3}\nem d^3t\,\,
			\BB{
				\pb{ H(t_1) }{ \pb{ H(t_2) }{ H(t_3) } }
				+
				\pb{ H(t_3) }{ \pb{ H(t_2) }{ H(t_1) } }
			}
\\
		&
		+
		\frac{1}{12}\mem
		\int_{t_1>t_2>t_3>t_4}\nem d^4t\,\,
			\left(\,\begin{aligned}[c]
				&
				\pb{ H(t_1) }{ \pb{ H(t_2) }{ \pb{ H(t_3) }{ H(t_4) } } }
				\\
				&+
				\pb{ H(t_2) }{ \pb{ H(t_3) }{ \pb{ H(t_4) }{ H(t_1) } } }
				\\
				&-
				\pb{ H(t_1) }{ \pb{ H(t_4) }{ \pb{ H(t_3) }{ H(t_2) } } }
				\\
				&-
				\pb{ H(t_4) }{ \pb{ H(t_3) }{ \pb{ H(t_2) }{ H(t_1) } } }
			\end{aligned}\,\right)
		+ \cdots
	\,.
\end{split}
\end{align}
Again, each integration variable here runs through the interval $[0,T]$.
Note a $\mathbb{Z}_2$ symmetry
that pairs up the terms in the parentheses,
which holds generically true at all higher orders:
$\pb{ H(t_1) }{ \pb{ H(t_2) }{ H(t_3) } }
\leftrightarrow
\pb{ H(t_3) }{ \pb{ H(t_2) }{ H(t_1) } }$,
$\pb{ H(t_2) }{ \pb{ H(t_3) }{ \pb{ H(t_4) }{ H(t_1) } } }
\leftrightarrow
\pb{ H(t_1) }{ \pb{ H(t_4) }{ \pb{ H(t_3) }{ H(t_2) } } }$,
etc.

Lastly,
it should be remarked that a diagrammatic representation of the Magnus series
is readily viable
via the Feynman rules of phase space sigma models
explicated in \rcite{wlf-ps}.
When implemented in terms of
Heaviside step functions,
the time ordering in the integral domains in \eqref{G+4}
turns into retarded propagators;
see \rcite{eikonal2}.

We end with a summary.
In \Sec{CURR},
the exponential generator $G$ was declared as the new currency
in classical mechanics,
boiling down 
the physicist mission of computing $U$
to the problem of computing $G$.
Accordingly,
in this section,
we have provided two concrete ways to compute the time-evolution symplectomorphism $U$:
the Dyson series in \eqref{Usol}
and the Magnus series in \eqref{Gsol}.
The former returns $U$ as a giant sum of differential operators,
whereas the latter returns the exponential generator $G$ as a scalar function.
Surely, the latter method will be adopted and spotlighted in this paper.
From a given $G$, the symplectomorphism $U$ follows by the exponentiation $U = \exp(X_G)$.

\newpage

\section{Computation of Scattering Symplectomorphism}

From the above discussions,
we have established phase-space-based formulations of classical mechanics
in which symplecticity is manifest.
In this section, we apply these frameworks to scattering problems,
from which the classical counterpart of the scattering matrix (S-matrix)
is concretely defined.

\subsection{Classical Scattering Theory}
\label{CST}

What is a scattering?
A scattering is a physical process in which a well-defined notion of 
\emph{in- and out-regions}
exists.
As a typical situation,
one can imagine a space (or spacetime)
where two empty regions sandwich an ``interacting zone.''
In the empty regions,
the particle moves freely,
meaning that we can solve its trajectories exactly.
However, in the interacting zone,
the particle is exerted by an extra force:
due to interactions with external fields, for instance.
In a scattering process,
the particle starts from one of the empty regions,
plunges into the interacting zone,
and escapes to the other empty region.
The empty region at the beginning is called the in-region;
the latter is called the out-region.
In the far past, $t \to -\infty$, the particle belongs to the in-region.
In the far future, $t \to +\infty$, the particle belongs to the out-region.

Mathematically, a scattering is modeled by the following Hamiltonian:
\begin{align}
	\label{Hsplit}
	H(t)
	\,=\,
		H^\circ 
		+ \e\mem H'(t)
	\,.
\end{align}
Here, $H^\circ$ governs the free motion of the particle,
taken as time-independent for simplicity.
$H'(t)$, on the other hand,
implements the extra force in the interacting zone,
taken as time-dependent for full generality.
These are called the free and interaction Hamiltonians, respectively.
Also, $\e$ is a (formal) parameter
introduced for describing how strongly does the particle couple
to the external fields in the interacting zone:
a sort of a \emph{coupling constant}.
For simplicity, we will treat $\e$ perturbatively.

By plugging in \eqref{Hsplit},
the Liouville equation in \eqref{Leq}
becomes
\begin{align}
	\label{SL.1}
	\dot{\rho}(t)
	-
		X_{H^\circ}[\rho(t)]
	\,=\,
		\e\mem
		X_{H'(t)}[\rho(t)]
	\,.
\end{align}
This describes the time evolution of a classical probability distribution
$\rho(t)$
that undergoes a scattering process.

To solve \eqref{SL.1},
we recall a familiar trick
from a typical chapter 
in standard quantum mechanics textbooks
\cite{griffiths2018introduction,shankar2012principles,sakurai2020modern}:
``Time-Dependent Perturbation Theory.''
Namely,
we take advantage of the assumption that
the free time-evolution symplectomorphism is known:
\begin{align}
	\label{U0}
	U^\circ(t_1,t_2)
	\,:=\,
		\exp\big(\hem{
			(t_1{\hem-\,}t_2)\mem X_{H^\circ}
		}\hhem\big)
	\,.
\end{align}
Using \eqref{U0},
we define the \emph{interaction picture}
such that \eqref{SL.1} boils down to
\begin{align}
	\label{Leq.int}
	\dot{\trho}(t)
	\,=\,
		\e\mem
		X_{\tH'(t)}[\trho(t)]
	\,.
\end{align}
Here, $\trho(t)$ and $\tH'(t)$ 
denote the classical probability distribution 
and the interaction Hamiltonian
in the interaction picture,
respectively.
It should be clarified that
the precise definition of the in-region and the out-region 
references $\tH'(t)$.
Namely,
the precise stipulation for a scattering problem is
\begin{align}
	\label{valid}
	\lim_{t\to \pm\infty}
		\tH'(t) 
		\,=\, 0
	\,.
\end{align}

\eqref{Leq.int}
is nothing but the Liovuille equation in the interaction picture,
which simply puts some tildes and primes 
on \eqref{Leq}.
Thus, from our discussion in \Sec{SOLVING},
the solution to \eqref{Leq.int} is immediate.
By referring to \eqref{U-Pexp},
the time-evolution symplectomorphism in the interaction picture
found as the ordered exponential,
\begin{align}
	\label{S-Pexp}
	S
	\mem=\,
		\Pexp{\,
			\e\mem
			\int_{-\infty}^{+\infty} dt\,\,
				X_{\tH'(t)}
		}
	\,\,\,:\,\,\,
		\trho(-\infty)
	\,\,\mapsto\,\,\hnem
		\trho(+\infty)
	\,.
\end{align}
By definition,
this $S$ describes the map from the initial phase space to the final phase space
in the interaction picture,
sending $\trho(-\infty)$ to $\trho(+\infty)$.
For obvious reasons,
we name this map the 
\emph{scattering symplectomorphism},
or \emph{S-symplectomorphism} \cite{ambikerr1} in short.

Again, the S-symplectomorphism can be 
computed 
via either the Dyson series
or the Magnus series.
From \eqref{Usol}, the Dyson series is given by
\begin{align}
\begin{split}
	\label{Usol.int}
	S
	\mem&=\,
		1
		+
		\e\mem
		\int_{-\infty}^{+\infty} dt_1\,\,
			X_{\tH'(t_1)}
		+
		\e^2\mem
		\int_{-\infty}^{+\infty} dt_1
		\int_{-\infty}^{t_1} dt_2\,\,
			X_{\tH'(t_1)}
			X_{\tH'(t_2)}
		+
		\cdots
	\,.
\end{split}
\end{align}
From \eqref{Gsol}, the Magnus series is given by
\begin{align}
	\label{Gsol.int}
	\chi
	\,=\,
		\e\mem
		\int_{-\infty}^{+\infty} dt\,\,
			\tH'(t)
		+
		\frac{\e^2}{2}\,
		\int_{-\infty}^{+\infty} dt_1
		\int_{-\infty}^{t_1} dt_2\,\,
			\pb{\tH'(t_1)}{\tH'(t_2)}
		+
		\cdots
	\,.
\end{align}
Here, we have denoted the exponential generator in the interaction picture
as $\chi$:
\begin{align}
    \label{exp.cl.int}
    S
        \mem=\,
    \exp(X_\chi)
    \,.
\end{align}
Here, the symplecticity of the S-symplectomorphism is explicitly ensured as
\begin{align}
	\label{expmapG.int}
	X_\chi \in \symp(\P,\omega)
	\qiq
	S \in \Symp(\P,\omega)
	\,.
\end{align}
We will call $\chi$ the \emph{scattering generator},
following \rcite{eikonal1}.
\eqrefs{exp.cl.int}{expmapG.int}
are the interaction picture equivalents of
\eqrefs{exp.cl}{expmapG}.

Amusingly, the scattering generator $\chi$
generates the total time evolution throughout the infinite span of time
within just ``one second,''\footnote{
	dimensionless unit time, to be precise
}
reproducing the effect of the entire history from the physical far past $t = -\infty$ to the physical far future $t = +\infty$!

This establishes the classical theory of scattering.
We emphasize that our exposition has been purely classical,
despite the exact parallel with quantum mechanics.
In particular, the S-symplectomorphism $S$
and the scattering generator $\chi$
have been defined purely within classical mechanics,
not by taking $\hbar\to0$ limits to quantum-mechanical expressions.

\subsection{Interaction Picture}
\label{INTPICTURE}

It remains to elaborate on the precise definition of interaction picture,
which we have swept under the rug
around \eqrefs{U0}{Leq.int}.

Let $t_*$ be a constant of one's choice.
For any time-dependent function $f(t)$ on the phase space,
such as $\rho(t)$ or $H'(t)$,
the interaction picture is defined as
\begin{align}
	\label{int-def}
	\tilde{f}(t)
	\,:=\,
		U^\circ(t_*,t)[ f(t) ]
	\,,
\end{align}
which means to drag/transport the function $f(t)$ \textit{backwards} in time
via the free dynamics.

It is not difficult to check that
the definition in \eqref{int-def}
implies
the differential equation in \eqref{Leq.int}
governing $\trho(t)$,
the classical probability distribution in the interaction picture.
This validates the fact that the Dyson series in \eqref{S-Pexp}
has indeed computed the $S$-symplectomorphism 
as the map sending 
$\trho(-\infty)$ to $\trho(+\infty)$:
\begin{align}
	\label{eve.int}
	S
	\,\,\,:\,\,\,
		\trho(-\infty)
	\,\,\mapsto\,\,
		\trho(+\infty)
	\,\qfq\,
	\trho(+\infty)
	\,=\,
		S[\mem
			\trho(-\infty) 
		\mem]
	\,.
\end{align}

To provide more details,
let $U(t_1,t_2)$ be the differential operator
representing the time-evolution
due to the full Hamiltonian $H(t)$ in \eqref{Hsplit}
from time $t_2$ to time $t_1$.
In the free theory limit $\e \to 0$,
this reduces to $U^\circ(t_1,t_2)$ in \eqref{U0}.

By definition, 
$U(+\infty,-\infty)$ is the map
sending $\rho(-\infty)$ to $\rho(+\infty)$:
\begin{align}
	\label{eve}
	U(+\infty,-\infty)
	\,\,\,:\,\,\,
		\rho(-\infty)
	\,\,\mapsto\,\,
		\rho(+\infty)
	\,\qfq\,
	\rho(+\infty)
	\,=\,
		U(+\infty,-\infty)[\mem
			\rho(-\infty) 
		\mem]
	\,.
\end{align}
With this understanding,
consider the following:
\begin{align}
\begin{split}
	\label{getS-1}
	\trho(+\infty)
	\mem&=\mem
		U^\circ(t_*,+\infty)[\mem
			\rho(+\infty)
		\mem]
	\,,\\
	\mem&=\mem
		U^\circ(t_*,+\infty)[\mem
		U(+\infty,-\infty)[\mem
			\rho(-\infty)
		\mem]
		\mem]
	\,,\\
	\mem&=\mem
		U^\circ(t_*,+\infty)[\mem
		U(+\infty,-\infty)\mem
		U^\circ(-\infty,t_*)[\mem
			\trho(-\infty)
		\mem]
		\mem]
	\,.
\end{split}
\end{align}
The first line describes that
$\trho(+\infty)$
is the classical probability distribution at the far future,
$\rho(+\infty)$,
dragged back to the designated time $t_*$.
The second line uses \eqref{eve}
to observe the full time evolution.
The third line uses \eqref{int-def} again,
returning to $t_*$.

Consequently, it follows from \eqrefs{eve.int}{getS-1} that
\begin{align}
	\label{Moller}
	S
	\,=\,
		U^\circ(t_*,+\infty)
		\circ
		U(+\infty,-\infty)
		\circ
		U^\circ(-\infty,t_*)
	\,,
\end{align}
where we understand that the right-hand side is defined by first regulating the infinities 
and then taking the limits.
This describes 
a ``loop''
based at a point $t_*$
on the time axis,
which visits first the far past ($-\infty$) and then the far future ($+\infty$).

This particular choreography in time,
drawing a loop based at $t_*$,
implies that
\eqref{Moller}
is the M{\o}ller operator
\cite{taylor2012scattering}
realized at time $t_*$,
implemented in classical mechanics.
Thus it indeed describes the classical incarnation of the S-matrix.

\subsection{Impulse of Classical Observables}

As an application of the S-symplectomorphism and scattering generator,
we provide a strictly classical derivation for the impulse of classical observables.

Given a classical probability distribution,
the expectation value of a classical observable $\O$
is given by \eqref{def-expval}.
When defined in the interaction picture, this is
\begin{align}
	\label{def-expval.int}
	\expval{\O}_t
	\,:=\,
		\int_\P\,\hem \O\mem \trho(t)\mem \mu
	\,.
\end{align}
As a result,
the impulse of $\O$
as an expectation value
is given by
\begin{align}
\begin{split}
	\label{imp-calc}
	\expval{\O}_{+\infty}
	-
	\expval{\O}_{-\infty}
	\,&=\,
		\int_\P\,\hem
			\O\mem
			\BB{
				\trho(+\infty) - \trho(-\infty)
			}\mem\mu
	\,,\\
	\,&=\,
		\int_\P\,\hem
			\O\mem
			\BB{
				S[\mem \trho(-\infty) \mem] - \trho(-\infty)
			}\mem\mu
	\,,\\
	\,&=\,
		\int_\P\,\hem
			\BB{
				S^{-1}[\mem \O \mem] - \O
			}\mem\trho(-\infty)
			\mem\mu
	\,=\,
		\expval{\Del\O}_{-\infty}
	\,,
\end{split}
\end{align}
where we have defined
\begin{align}
	\label{imp-def}
	\Del\O
	\,:=\,
		S^{-1}[\mem \O \mem] - \O
	\,.
\end{align}
Here, $S^{-1}$ acts on $\O$ 
as the differential operator
inverse to $S$.
The third equality in \eqref{imp-calc}
uses the invariance of the Liouville measure $\mu$
while performing integration by parts.

\eqref{imp-calc} implies that
the impulse arises as the expectation value
of $\Del\O$ 
computed at the initial time.
Since this holds for any initial distribution $\trho(-\infty)$,
we can simply identify $\Del\O$
as the impulse of $\O$.
Expanding $S^{-1}$ in the exponential representation, we find
\begin{align}
\begin{split}
	\label{expKMOC}
	\Del\O
	\,&=\,
		\mathe^{\pb{\blank}{\chi}}[\mem \O \mem] - \O
	\,,\\
	\,&=\,
		\pb{\O}{\chi}
		+ \frac{1}{2!}\,
			\pb{\pb{\O}{\chi}}{\chi}
		+ \frac{1}{3!}\,
			\pb{\pb{\pb{\O}{\chi}}{\chi}}{\chi}
		+ \cdots
	\,.
\end{split}
\end{align}
Of course, one could also directly compute $S^{-1}[\mem \O \mem]$
by inverting the Dyson series in \eqref{Usol.int}.
Yet,
employing the exponentiated form as in \eqref{expKMOC}
nicely organizes the formula in terms of nested Poisson brackets,
which uses the Magnus series formula in \eqref{Gsol.int}.

In the literature, the Kosower-Maybee-O'Connell formula \cite{KMOC}
has been employed to obtain
the classical impulse from the S-matrix,
where one checks the cancellation of superclassical terms.
The above derivation
provides an approach strictly defined within classical mechanics,
obtaining the classical impulse from the S-symplectomorphism
(with or without the exponential parametrization).

We end with a summary:
\begin{align}
\begin{split}
    {\renewcommand{\arraystretch}{1.1}
    \begin{array}{rl}
        \text{
            S-symplectomorphism, $S$
        }
        &\,\,\,\xleftrightarrow[\,]{}\,\,\,
        \text{
            S-matrix, $\hat{S}$
        }
        \,,\\
        \text{
            Scattering Generator, $\chi$
        }
        &\,\,\,\xleftrightarrow[\,]{}\,\,\,
        \text{
            Eikonal Matrix, $\hat{\chi}$
        }
        \,.
    \end{array}}
\end{split}
\end{align}
Here, the eikonal matrix
in quantum mechanics
is defined as
\begin{align}
	\label{S-chi-quantum}
	\hat{S}
	\,=\,
		\exp\bb{{
			\frac{1}{i\hbar}\mem
				\hat{\chi}
		}}
	\qfq
	\hat{\chi}
	\,=\,
		i\hbar\, \log\hat{S}
	\,.
\end{align}
Accordingly, the scattering generator $\chi$ has been also called the classical eikonal \cite{eikonal2}.
The interpretation of $\hat{\chi}$ as an effective Hamiltonian for a unit-time evolution
should be evident from \eqref{S-chi-quantum}.

The astute reader 
will uncover
more parallels
we have employed so far.
The time-evolution differential operator
$U(t_1,t_2)$
has paralleled
the time-evolution unitary operator
$\hat{U}(t_1,t_2)$
in quantum mechanics,
while
the classical probability distribution 
$\rho(t)$
has been an analogy for
the quantum density matrix
$\hat{\rho}(t)$.
It should be clear that,
based on such similarities,
the S-symplectomorphism $S$
and the scattering generator $\chi$
can also be derived by
taking $\hbar \to 0$ limits
to appropriate quantum-mechanical equations:
see \rrcite{eikonal2,eikonal5}.

We point out that promoting
these analogies
to strict correspondences
is viable
by employing the phase space formulation of quantum mechanics
\cite{Moyal:1949sk,Groenewold:1946kp,Wigner:1932eb,weyl1927quantenmechanik}:
see \App{DEFQUANT}.

\section{Concrete Examples}
\label{EXAMPLES}

Eventually, we provide more explicit examples 
to assist a concrete understanding.

\subsection{Falling Apple in Space Elevator}

Suppose our falling apple in \Sec{CURREX}
is now put on a ``space elevator'':
an elevator installed in outer space,
moving along a one-dimensional axis
with a custom time-dependent acceleration.
Let $\a(t)\mem g$ be the resulting effective gravitational acceleration
in the elevator's frame.
Here, $g$ is a constant while $\a(t)$ is dimensionless and time-dependent.

In this case, we have the Hamiltonian system
\begin{align}
	\label{apple+}
	\omega \,=\,
		dp \wedge dx
	\,,\quad
	\pb{x}{p} \,=\,
		1
	\,,\quad
	H(x,p;t) \,=\,
		\frac{p^2}{2m} + mgx\, \a(t)
	\,,
\end{align}
where we have started to spell out all arguments of phase space functions explicitly.
The equations of motion are
\begin{align}
\begin{split}
    \label{eom-apple+}
    \dot{x}(t) \,=\, p(t)/m
    \,,\quad
    \dot{p}(t) \,=\, - mg\, \a(t)
    \,.
\end{split}
\end{align}

In particular, let us suppose $\a(t)$ is a smooth function
which asymptotes to zero when $t\to\pm\infty$.
What is the time evolution of the apple
from $t = -\infty$ to $t = +\infty$?
To this end, we view this system as a scattering problem
in the template of \eqref{Hsplit}
with $g$ as the coupling constant:
\begin{align}
	\label{Hsplit.apple+}
	H^\circ(x,p)
	\,=\,
		\frac{p^2}{2m}
	\,,\quad
	H'(x,p;t)
	\,=\,
		mx\,\a(t)
	\,.
\end{align}

\subsubsection{Interaction Picture $=$ Insert Free Trajectory}

The free time evolution
as a differential operator
in \eqref{U0} is given by
\begin{align}
	\label{U0straight}
	U^\circ(t_1,t_2)
	\,=\,
		\exp\BB{\hnem
			(t_1{\hem-\,}t_2)\mem
			\pb{H^\circ}{\blank}
		}
	\,=\,
		\exp\bb{
			-(t_1{\hem-\,}t_2)\mem
				\frac{p}{m}\, \frac{\partial}{\partial x}
		}
	\,,
\end{align}
which simply implements straight-line trajectories:
constant-velocity drifts,
$(x,p) \mapsto (x - (t_1-t_2)\mem p/m,p)$.
Consequently, the interaction picture
is given by
\begin{align}
	\label{int.apple+}
	\tilde{f}(x,p;t)
	\,=\,
		f\bigbig{x{\,+\,}pt/m,p;t}
	\,,
\end{align}
according to the definition in \eqref{int-def}
that puts $t_1 = t_*$ and $t_2 = t$ in \eqref{U0straight}.
For simplicity, we have set $t_* = 0$ as well.
Notably, \eqref{int.apple+} plugs in the \textit{forward}-in-time free time evolution
in the arguments of the function $f(x,p;t)$:
$(x,p) \mapsto (x+pt/m,p)$.

Earlier, we have remarked that
\eqref{int-def} transports the function \textit{backwards} in time.
No confusion should arise from this apparent flip:
this simply describes the usual wisdom that
mapping $f(x)$ to $f(x {\,+\,} a)$
translates the function by $-a$ instead of $+a$.

Therefore, we conclude that 
the interaction picture 
in classical mechanics
simply means
to plug in the free trajectory to a function,
evolved forward in time.
To repeat,
\begin{align}
	\label{CIP}
	\text{Interaction Picture}
	\,=\,
	\text{Insert Free Trajectory
	}.
\end{align}
In fact,
the statement in \eqref{CIP}
has been established by
Campbell {et al.}\:\cite{Campbell:1975nn}
(see ``III. Classical Interaction Picture'').
A recent revival has taken place in \rcite{eikonal2}.\footnote{
	See also \rcite{eikonal5}.
	In the meantime, \rcite{ambikerr1}
	pursued a Dirac bracket approach 
	to implement the interaction picture
	in relativistic scattering.
}
In this paper,
we have provided a consistent derivation
in the differential operator language.

With this understanding,
the interaction Hamiltonian in \eqref{Hsplit.apple+}
is described in the interaction picture as
\begin{align}
	\label{tH.apple+}
	\tH'(x,p;t)
	\,=\,
	H'(x{\,+\,}pt/m,p;t)
	\,=\,
		\BB{mx+pt}\, \a(t)
	\,.
\end{align}
On account of \eqref{valid},
the scattering problem is valid if 
$t\mem \a(t) \to 0$
as $t\to\pm\infty$.

\subsubsection{Scattering Generator}

Now we can readily apply the Magnus series in \eqref{Gsol.int}
to find the scattering generator.
From \eqref{tH.apple+}, it follows that
\begin{align}
	\pb{\tH'(x,p;t_1)}{\tH'(x,p;t_2)}
	\,=\,
		- 
		m\hem
		(t_1{\mem-\,}t_2)
		\,\a(t_1)\,\a(t_2)
	\,,
\end{align}
while higher nested brackets vanish.
Therefore, the Magnus series in \eqref{Gsol.int} gives
\begin{align}
	\label{G.apple+}
	\chi(x,p)
	\,=\,
		g
		\int dt_1\,\,
			\BB{\hnem
				mx + pt_1
			}\, \a(t_1)
		- \frac{1}{2}\mem mg^2\,
		\int_{t_1>t_2} d^2t\,\,
	 		(t_1{\mem-\,}t_2)
	 		\,\a(t_1)\,\a(t_2)
	 \,,
\end{align}
where $t_1,t_2 \in (-\infty,+\infty)$.
The second term in \eqref{G.apple+} simply describes
a constant function over the phase space,
but we are keeping it for the exact matching later.

According to the nested bracket calculation in \eqref{expKMOC},
the scattering generator in \eqref{G.apple+}
transforms the position $x$ and momentum $p$
(as classical observables)
as
\begin{align}
	\label{map.apple+}
	x
	\,\,\mapsto\,\,
		x + g \int_{-\infty}^{+\infty} dt_1\, t_1\mem \a(t_1)
	\,,\quad
	p
	\,\,\mapsto\,\,
		p - mg \int_{-\infty}^{+\infty} dt_1\, \a(t_1)
	\,.
\end{align}
Indeed, it is not difficult to reproduce \eqref{map.apple+}
by performing the ``choreography'' by hand:
inertial motion from $t=0$ to $t=-T$,
accelerated motion till $t=T$,
then inertial motion back to $t=0$,
where $T$ regulates the infinity.
The first integral in \eqref{map.apple+}
will arise as
\begin{align}
	T
	\int_{-T}^{+T}\nem dt_1\,\,
		\a(t_1)
	- 
	\int_{-T}^{+T}\nem dt_1
	\int_{-T}^{t_1}\nem dt_2\,\,
		\a(t_2)
	\,=\,
	\int_{-T}^{+T}\nem dt_1\,\,
		t_1\mem \a(t_1)
	\,.
\end{align}
\phantom{.}\vspace{-0.7\baselineskip}

\subsubsection{On-Shell Action}

Meanwhile, let us see how \eqref{map.apple+}
is reproduced from the Hamilton-Jacobi formalism.
To this end, we compute the on-shell action in the interaction picture.
In general, the Lagrangian reproducing
the equations of motion in the interaction picture
is \cite{Campbell:1975nn}
\begin{align}
	\label{intL}
		p \dot{x}
		- \e\mem \tH'(x,p;t)
	\,,
\end{align}
which takes $\e\mem \tH'(x,p;t)$ as the Hamiltonian.

When applied to
our specific example of the falling apple,
the second principal function in the interaction picture
is given by
\begin{align}
	I^{(2)}(p',x)
	\,=\,
	\Extr{
		p(+\infty) = p'\\
		x(-\infty) = x
	}{
		-p(+\infty)\, x(+\infty)
		+ \int_{-\infty}^{+\infty} dt\,\,
			\bb{
				p(t)\mem \dot{x}(t)
				- g\mem \BB{mx+pt}\mem \a(t)
			}
	}
	\,.
\end{align}
Here, we have used $I$ instead of the previous symbol $F$
to explicitly remind the viewer that 
this on-shell action is defined and computed
in the interaction picture.
The extremum is achieved by the saddle
\begin{align}
	\label{saddle.apple+}
	x(t)
	\,=\,
		x + g \int_{-\infty}^t dt_1\,\,
			t_1\mem \a(t_1)
	\,,\quad
	p(t)
	\,=\,
		p' + mg \int_t^{+\infty} dt_1\,\, \a(t_1)
	\,,
\end{align}
from which it follows that
\begin{align}
	\label{F2.apple+}
\begin{split}
	I^{(2)}(p',x)
	\,=\,
	-
\left(\,
\begin{aligned}[c]
	&
		{p'x}
		+gp'
		\int_{-\infty}^{+\infty}\nem dt_1\,\,
			t_1\mem \a(t_1)
		+mgx
		\int_{-\infty}^{+\infty}\nem dt_1\,\,
			\a(t_1)
	\\
	&
		+mg^2
		\int_{-\infty}^{+\infty}\nem dt_1
			\int_{-\infty}^{t_1} dt_2\,\,
				t_2\mem \a(t_1)\mem \a(t_2)
\end{aligned}
\hem\right)
	\,.
\end{split}
\end{align}

From the first-derivatives of \eqref{F2.apple+},
we find
\begin{align}
\begin{split}
	\label{HJ.apple+}
	x'
	\,&=\,
		-\frac{\partial}{\partial p'}\,
			I^{(2)}(p',x)
	\,=\,
		x + g
			\int_{-\infty}^{+\infty}\nem dt_1\,\,
				t_1\mem \a(t_1)
	\,,\\
	p
	\,&=\,
		-\frac{\partial}{\partial x}\,
			I^{(2)}(p',x)
	\,=\,
		p' + mg
		\int_{-\infty}^{+\infty}\nem dt_1\,\,
			\a(t_1)
	\,,
\end{split}
\end{align}
due to the differential relation stated in \eqref{dF}.
Clearly, \eqref{HJ.apple+} reproduces \eqref{map.apple+}.

\subsubsection{Matching in the Interaction Picture}

Do the scattering generator in \eqref{G.apple+}
and the on-shell action in \eqref{F2.apple+}
take the same value?
Even when re-expressed in terms of the initial variables $x$ and $p$,
the on-shell action 
in \eqref{F2.apple+}
contains a term quadratic in phase space coordinates ($-px$),
while the scattering generator in \eqref{G.apple+}
is at most linear in phase space coordinates.

One might consider dropping the boundary term $p'x$.
However, the interaction-picture saddle in 
\eqref{saddle.apple+}
seems to imply that
fixing positions at both ends
does not serve as a sensible boundary condition.

Instead, we try the matching condition 
proposed in \eqref{matching}.
In the interaction picture,
it becomes
\begin{align}
\begin{split}
	\label{matching.int}
	&
	\underset{\substack{
		x(+\infty) = x'\\
		x(-\infty) = x
	}}{\Ext}\,
	\bbsq{
    \int_{-\infty}^{+\infty} dt\,\,
    	\BB{
	    	p(t)\mem \dot{x}(t) {\,-\,} \e\mem \tH'(x(t),p(t);t)
    	}
    }
    \\
   	&
	=\,
	\underset{\substack{
		x(1) = x'\\
		x(0) = x
	}}{\Ext}\,
	\bbsq{
    \int_{0}^{1} dt\,\,
    	\BB{
	    	p(t)\mem \dot{x}(t) {\,-\,} \chi(x(t),p(t))
    	}
	}
    \,,
\end{split}
\end{align}
where one is free to add boundary terms 
as commented earlier.
The quantum-mechanical argument provided in 
\Sec{MATCHING}
readily applies
through proper implementations of the interaction picture,
following \rcite{Campbell:1975nn}.

For the mixed (position-momentum) boundary condition,
this matching condition is
\begin{align}
	\label{matching.apple+}
	I^{(2)}(p',x)
	\,=\,
	\Extr{
		p(1) = p'\\
		x(0) = x
	}{
		-p(1)\mem x(1)
		+ \int_0^1 dt\,\,
			\bb{
				p(t)\mem \dot{x}(t)
				- \chi(x(t),p(t))
			\hnem}
	}
	\,,
\end{align}
where the left-hand side refers to
the original computation 
in the Hamilton-Jacobi formalism.
The right-hand side,
on the other hand,
derives an on-shell action
from $\chi(x,p)$ as a time-independent Hamiltonian.

Since $\chi(x,p)$ is time-independent and also at most linear in $x,p$,
the saddle for \eqref{matching.apple+}
simply exhibits linear dependency in time:
\begin{align}
\begin{split}
	\label{lineartrj}
	x(t)
	\,&=\,
		x+
		gt
		\int_{-\infty}^{+\infty} dt_1\,\,
			t_1\mem \a(t_1)
	\,,\quad
	p(t)
	\,=\,
		p'
		+mg(1{\mem-\,}t)
		\int_{-\infty}^{+\infty} dt_1\,\,
			\a(t_1)
	\,.
\end{split}
\end{align}
Plugging in \eqref{lineartrj},
the right-hand side of \eqref{matching.apple+}
evaluates to
\begin{align}
	\label{F2re.apple+}
\begin{split}
	-
\left(\,
\begin{aligned}[c]
	&
		{p'x}
		+gp'
		\int_{-\infty}^{+\infty}\nem dt_1\,\,
			t_1\mem \a(t_1)
		+mgx
		\int_{-\infty}^{+\infty}\nem dt_1\,\,
			\a(t_1)
	\\
	&
		+
		\frac{1}{2}\mem mg^2\mem
			\bb{
				\int_{-\infty}^{+\infty} dt_1\,\,
						\a(t_1)
			}\bb{
				\int_{-\infty}^{+\infty} dt_2\,\,
						t_2\mem \a(t_2)
			}
	\\
	&
		+
		\frac{1}{2}\mem mg^2\,
		\int_{t_1>t_2} d^2t\,\,
	 		(t_1{\mem-\,}t_2)
	 		\,\a(t_1)\,\a(t_2)
\end{aligned}
\hem\right)
	\,.
\end{split}
\end{align}
It is easy to show that \eqref{F2re.apple+}
exactly reproduces 
\eqref{F2.apple+}.

\subsubsection{Matching in the Bulk}

To clarify,
it should be remarked that
this matching relation
is orthogonal to both the interaction picture and the scattering context.
That is, the matching relation
always works perfectly.

To demonstrate this point
explicitly,
let us assume a linear time profile for the acceleration
for simplicity: $\a(t) = \a\mem t$,
where $\a$ is a constant.
Let the initial and final times be
$\ti$ and $\tf = \ti + T$.
In this case,
we use the entire Hamiltonian
$H(x,p;t) = p^2/2m + mgx + m\a t x$
without the free-interaction split.
Straightforward computation shows that
the exponential generator, describing the ``bulk-to-bulk propagation,'' is
\begin{align}
	\label{lina-G}
	G(x,p)
	\,=\,
	T\,\bb{
		\frac{p^2}{2m}
		+ mx\mem\BB{
			g + \a \ti
		}
	}
	+
		\frac{T^2}{2}\mem m\a x
	+
		\frac{T^3}{12}\mem p \a 
	+ 
		\frac{T^5}{240}\mem m \a^2
	\,,
\end{align}
whereas the on-shell action,
computed either from the original description due to $H(x,p;t)$
or the effective description due to $G(x,p)$ in \eqref{lina-G},
is given by
\begin{align}
	\label{lina-F}
\begin{split}
	F^{(2)}(p',x)
	\,=\,
	-
\left(\,
\begin{aligned}[c]
	&
		{p'x}
		+
		T\,\bb{
			\frac{p^2}{2m}
			+ m\mem\BB{g+\a\ti}\mem x
		}
	\\
	&
		+
		\frac{T^2}{2}\,\BB{
			g p'
			+ p' \a \ti
			+ m \a x
		}
		+
		\frac{T^3}{6}\,\BB{
			2p'\a
			+ mg
			+ m \a \ti
		}
	\\
	&
		+
		\frac{5T^4}{24}\,
			m\a\mem\BB{g+\a\ti}
		+
		\frac{T^5}{15}\,
			m\a^2
\end{aligned}
\hem\right)
	\,.
\end{split}
\end{align}

\subsection{Anharmonic Oscillator}

For the next example, 
consider an oscillator.
The free theory is given by
\begin{align}
	\label{sho}
	\omega \,=\,
		dp \wedge dx
	\,,\quad
	\pb{x}{p} \,=\,
		1
	\,,\quad
	H^\circ(x,p;t) \,=\,
		\frac{p^2}{2m} + \frac{1}{2}\mem m\omega^2\mem x^2
	\,.
\end{align}
A nicer coordinate system for this phase space is
\begin{align}
	a \,=\,
		\frac{1}{\sqrt{2m\omega}}\,
			\BB{m\omega x + ip}
	\,,
\end{align}
which identifies $\R^2$
as a complex plane $\C$.
Then \eqref{sho} boils down to
\begin{align}
	\label{sho.aa}
	\omega \,=\,
		i\mem d\ba \wedge da
	\,,\quad
	\pb{a}{\ba} \,=\,
		-i
	\,,\quad
	H^\circ(a,\ba;t) \,=\,
		\omega\mem \ba a
	\,.
\end{align}
The free evolution is given by
\begin{align}
	a
	\,\,\mapsto\,\,
	a\mem \mathe^{-i\omega t}
	\,,\quad
	\ba
	\,\,\mapsto\,\,
	\ba\mem \mathe^{+i\omega t}
	\,.
\end{align}

So far, our interaction Hamiltonians have been at most linear in phase space coordinates.
To probe the matching relation at further nonlinear orders,
we take
\begin{align}
	\label{intHaa}
	H'(a,\ba)
	\,=\,
		g\,\BB{
			a^2 + \lambda\mem \ba
		}
	\qiq
	\tH'(a,\ba;t)
	\,=\,
		g\,\BB{
			a^2\mem \mathe^{-2i\omega t} + \lambda\mem \ba\mem \mathe^{+i\omega t}
		}
	\,,
\end{align}
where $g,\lambda$ are constants.
This provides a nice toy example 
that activates the Magnus series in \eqref{Gsol}
up to the third order,
despite a formal complexification.
We have explicitly checked by computer algebra
that the matching condition holds for the (complexified) on-shell action
\begin{align}
	\label{Faa}
	F(\ba',a)
	\,=\,
	\Extr{
		\ba(T) = \ba'\\
		a(0) = a
	}{
		-i\mem \ba(T)\mem a(T)
		+ \int_0^T dt\,\,
			\bb{
				i\mem \ba(t)\mem \dot{a}(t)
				- G(a(t),\ba(t))
			\hnem}
	}
	\,,
\end{align}
though its explicit value is rather uninsightful to spell out.
Similarly, we have also imposed other nonlinear forms of the interaction Hamiltonian in \eqref{intHaa}
and verified match.

Another interesting oscillator example one can think about
is driven oscillator,
in which case one can carry out a classical computation 
suggestive of the Fermi golden rule.

On a related note, it should be interesting to study adiabatic invariants
and action-angle coordinates
from the current frameworks:
a mass on a ``rusting'' spring, for instance.

\subsection{Spin in Magnetic Field}
\label{MAGSPIN}

Eventually,
we provide an extreme example
in which the on-shell action is not even defined,
although the exponential generator is perfectly well-defined.
This sharply illuminates a fundamental distinction
between the on-shell action and the exponential generator.

So far,
we have exclusively presumed symplectic manifolds for the mathematical model for phase space.
However,
strictly speaking,
the minimal mathematical setup for a phase space
is a \emph{Poisson manifold}:
a smooth manifold equipped with a Jacobi-satisfying Poisson bracket.
A symplectic manifold is a Poisson manifold.
But crucially, a Poisson manifold is not necessarily a symplectic manifold.
This is related to the fact that the Poisson bracket
describes the inverse of the symplectic form,
as is remarked in
\Sec{SYMP}.

A physically relevant example of a Poisson manifold
is the space $\R^3$ equipped with coordinates
$(J_1,J_2,J_3)$ and the Poisson brackets
\begin{align}
	\label{Jpb}
	\pb{J_i}{J_j}
	\,=\,
		\ve^k{}_{ij}\mem J_k
	\,,
\end{align}
which is sometimes called the isotropic Poisson-noncommutative three-space.
This is nothing but a phase space for angular momentum.
Evidently, this phase space is odd-dimensional,
so it cannot be a symplectic manifold on its own.

To reiterate, there is no symplectic form admitted in this space.
In turn, there is no symplectic potential.
Therefore, one cannot formulate an action principle
nor a principal function.
As a result, the Hamilton-Jacobi formalism is not applicable in this phase space.

However, 
the exponential generator 
takes only the Poisson brackets 
and the equations of motion
as strict requirements.
In a Poisson manifold, the vector field $X_f$ is defined by \eqref{Xpb}
by taking the Poisson bracket as granted.
It then follows that the Jacobi identity of the Poisson bracket ensures \eqref{X-repr}.
Time evolution is now a \emph{Poisson diffeomorphism} $U$,
which can be written as the exponentiation 
$U = \exp(X_G)$
of a \emph{Poisson vector field} $X_G$.

For instance, take the time-dependent Hamiltonian 
\begin{align}
	\label{HJ}
	H(\vec{J};t)
	\,=\,
		\gamma\mem \vec{B}(t) \cdot \vec{J}
	\,,
\end{align}
where we have employed the three-vector notation:
$\vec{J} = (J_1,J_2,J_3)$.
This Hamiltonian can describe 
a spinning top or 
a spinning particle 
in an external time-dependent magnetic field $\vec{B}(t)$.
The Hamiltonian equations of motion are given by
\begin{align}
	\label{eomJ}
	\dot{\vec{J}}(t)
	\,=\,
		\gamma\mem \vec{B}(t) \mtimes \vec{J}(t)
	\,,
\end{align}
which follows by computing $\pb{\vec{J}}{H(\vec{J};t)}$.
Physically, this describes spin precession
due to the external magnetic field.
The solution to \eqref{eomJ} is, surely,
\begin{align}
	\vec{J}(T)
	\,=\,
		\Pexp{\,
			\gamma
			\int_{0}^{T} dt\,\,
				\vec{B}(t) \times
		}\,
		\vec{J}(0)
	\,,
\end{align}
from which a rotation matrix---%
an $\mathrm{SO}(3)$ group element---%
is identified as
\begin{align}
	\label{dyson-R}
	R
	\,=\,
		\Pexp{\,
			\gamma
			\int_{0}^{T} dt\,\,
				\vec{B}(t) \times
		}
	\,\,\,\in\,\,\,
	\mathrm{SO}(3)
	\,.
\end{align}
From the Magnus series,
the total angle of rotation
is found as a pseudovector
\begin{align}
	\label{magnus-R}
	\vec{\Theta}
	\,=\,
		\gamma
		\int_0^T dt_1\,\,
			\vec{B}(t_1)
		+
		\frac{\gamma^2}{2}\,
		\int_0^T dt_1
		\int_0^{t_1} dt_2\,\,
			\vec{B}(t_1) \times \vec{B}(t_2)
		+ \cdots
	\,,
\end{align}
so that
$
	R
	=
	\exp(\vec{\Theta}\hem{\times}\nem)
$.
Mathematically, this describes
the exponentiation of the Lie algebra $\mathrm{so}(3)$
to
the Lie group $\mathrm{SO}(3)$.
The Lie bracket for $\mathrm{so}(3)$
is simply the cross product.

Similarly, 
the Liouville equation is given by
\begin{align}
	\label{liouvJ}
	X_{H(t)}
	\,=\,
		\gamma\mem
			(\vec{J} \mtimes \vec{B}(t))
			\cdot
			\frac{\partial}{\partial \vec{J}}
	\qiq
	\dot{\rho}(\vec{J};t)
	\,=\,
		\gamma\mem
			(\vec{J} \mtimes \vec{B}(t))
			\cdot
			\frac{\partial}{\partial \vec{J}}
			\,\,
			\rho(\vec{J};t)
	\,.
\end{align}
By solving the Liouville equation,
the differential operator $U$
representing $R$
is read off as
\begin{align}
	\label{magnus-RU}
	U
	\,=\,
		\exp\nem\bigg(\,{
				(\vec{J} \mtimes \vec{\Theta})
				\cdot
				\frac{\partial}{\partial \vec{J}}
		\,}\mem\bigg)
	\qiq
	\rho(\vec{J};T)
	\,=\,
		U[\mem{
			\rho(\vec{J};0)
		}\mem]
	\,.
\end{align}
Finally,
the exponential generator $G$
that generates the total rotation
in just ``one second'' is given by
\begin{align}
	\label{GJ}
	G
	\,=\,
		\vec{J} \cdot \vec{\Theta}
	\,=\,
		\vec{J}
		\cdot
		\bb{
			\gamma
			\int_0^T dt_1\,\,
				\vec{B}(t_1)
			+
			\frac{\gamma^2}{2}\,
			\int_0^T dt_1
			\int_0^{t_1} dt_2\,\,
				\vec{B}(t_1) \times \vec{B}(t_2)
			+ \cdots
		}
	\,,
\end{align}
such that
\begin{align}
	U
	\,=\,
		\exp(X_G)
	\,=\,
		\exp\nem\bigg(\,{
				\vec{J} \mtimes \frac{\partial G}{\partial \vec{J}}
				\cdot
				\frac{\partial}{\partial \vec{J}}
		\,}\mem\bigg)
	\,.
\end{align}
It is left as an exercise to reproduce \eqref{GJ}
by plugging in \eqref{HJ} to \eqref{Gsol}.
Note the identity
$
	\pb{
		\vec{J}\mdot \vec{\alpha}
	}{
		\vec{J}\mdot \vec{\beta}
	}
	=
		\vec{J} \cdot (\vec{\alpha}\mtimes\vec{\beta})
$.

From this physical example,
it is made clear that
the exponential generator $G$
is perfectly well-defined even in Poisson manifolds.
The on-shell action $F$,
however,
cannot be defined
due to the absence of a Lagrangian.

Still, one may insist a Lagrangian formulation.
As explicated in \Sec{SYMP},
an action principle is always viable
in a symplectic manifold
via the symplectic potential in local patches.
Hence one considers symplectic realizations of the above Poisson manifold $\R^3$,
which means to implement the angular momentum $\vec{J}$ as a set of composite variables
arising from more microscopic degrees of freedom.
Note that
the Poisson manifold $\R^3$ has taken $\vec{J}$
directly as its coordinates,
thus viewing $\vec{J}$ as fundamental degrees of freedom.

Concretely, one can take either
the well-known Schwinger oscillator model
\cite{jordan1935zusammenhang,schwinger1952angular,sakurai2020modern}
or the Poisson-noncommutative sphere $S^2 \cong \mathbb{CP}^1$
\cite{varilly1989moyal,shankar2012principles}.
This facilitates writing down an action and,
in turn,
the Hamilton-Jacobi formalism.
It can be checked that our matching relation works
in such symplectic realizations.

\medskip
\noindent
\textbf{Acknowledgements.}
As a general comment,
the author wishes to thank his collaborators,
Jung-Wook Kim,
Sungsoo Kim,
and
Sangmin Lee,
for continued discussions on the scattering generator and Magnus series.
J.-H.K. is supported by the Department of Energy (Grant {No.}~DE-SC0011632) and by the Walter Burke Institute for Theoretical Physics.

\appendix

\section{S-matrix $=$ Noncommutative S-symplectomorphism}
\label{DEFQUANT}

In the phase space formulation of quantum mechanics \cite{Moyal:1949sk,Groenewold:1946kp,Wigner:1932eb,weyl1927quantenmechanik},
also known as deformation quantization,
one deforms the ring of functions
on the phase space $(\P,\omega)$
by replacing the product $(f,g) \mapsto fg$
to an associative yet noncommutative product $(f,g) \mapsto f\star g$.
This makes the phase space
a noncommutative geometry
\cite{connes1993non}.
The noncommutative product $\star$,
referred to as the \emph{star product},
can be given in any Poisson manifolds
\cite{kontsevich}.
For physicists,
the star product
is simply an avatar for the quantum operator product
with a specified ordering prescription.
In particular, the symmetric (Weyl) ordering 
corresponds to the Moyal \cite{Moyal:1949sk,Groenewold:1946kp}
star product.
An exact map between 
the operator product in the standard formulation
and the star product in the phase space formulation
can be established via Stratonovich-Weyl kernels \cite{weyl1927quantenmechanik,stratonovich1957distributions,Varilly:1989sv}.

The axioms of the star product are
\begin{align}
\label{star-axioms}
	1 \star f \,=\, f \star 1 \,&=\, f
	\,,\quad
	f \star g 
	\,=\,
		fg + \O(\hbar^1)
	\,,\quad
	f \star g - g \star f
	\,=\,
		i\hbar\, \pb{f}{g}
		+ \O(\hbar^2)
	\,.
\end{align}
For each function $f$ in phase space,
there is a unique associated differential operator $X_f^\star$:
\begin{align}
	\label{Xpb.star}
	X_f^\star[g]
	\,=\,
		\pb{f}{g}_\star
	\,.
\end{align}
Here, $\pb{f}{g}_\star$ is the \emph{deformed Poisson bracket}, defined as
\begin{align}
	\label{pb.star}
	\pb{f}{g}_\star
	\,:=\,
		\frac{1}{i\hbar}\,
			\BB{
				f \star g
				-
				g \star f
			}
	\,.
\end{align}
An important identity reads
\begin{align}
    \label{X-repr.star}
    [ X_f^\star , X_g^\star ] \,=\, X_{\pb{f}{g}_\star}^\star
    \,.
\end{align}
The exponentiation of $X_f^\star$ returns a differential operator,
which we identify as describing a \emph{noncommutative symplectomorphism}.
The group of noncommutative symplectomorphisms is denoted as $\Symp_\star(\P,\omega)$.

By definition, the star product describes the quantum operator product:
$f \star g \leftrightarrow \hat{f}\hat{g}$.
Hence the deformed Poisson bracket in \eqref{pb.star} describes
the quantum commutator $[\hat{f},\hat{g}]$, divided by $i\hbar$.

In the standard formulation of quantum mechanics,
operators are elements of $\GL(\mathcal{H})$, where $\mathcal{H}$ is a Hilbert space.
In the phase space formulation of quantum mechanics,
operators are just functions on the phase space.
In particular, the density matrix $\hat{\rho}(t)$
corresponds to a quasiprobability distribution $\rho(t)$
on phase space,
governed by the $\hbar$-deformed version of the Liouville equation:
\begin{align}
	\label{Leq.star}
	\dot{\hat{\rho}}(t)
	\,=\,
		\frac{1}{i\hbar}\,
			[ \hat{H}(t) , \hat{\rho}(t) ]
	\quad\leftrightarrow\quad
	\dot{\rho}(t)
	\,=\,
		X_{H(t)}^\star[\rho(t)]
	\,.
\end{align}
As a result,
the quantum time evolution is a noncommutative symplectomorphism
mapping the initial $\hbar$-deformed phase space to the final $\hbar$-deformed phase space:
\begin{align}
	\label{U-cl.star}
    U
        \,\,\in\,\,
    \Symp_\star(\P,\omega)
    \,.
\end{align}
Concretely, this is defined from the solution to \eqref{Leq.star}.
The Dyson series approach
yields a mere transcription of the well-known quantum mechanical formula
to deformation quantization,
which describes star-exponentials.
On the other hand, the Magnus series yields \eqref{G+4}
with every Poisson bracket
replaced with the deformed Poisson bracket in \eqref{pb.star}:
\begin{align}
	\label{Gsol.star}
	G
	\,=\,
		&
		\int_0^T dt\,\,
			H(t)
		+
		\frac{1}{2}\,
		\int_0^T dt_1
		\int_0^{t_1} dt_2\,\,
			\pb{H(t_1)}{H(t_2)}_\star
		+
		\cdots
	\,.
\end{align}
This is by virtue of the correspondence in \eqref{X-repr.star}.

In the vanishing noncommutativity limit, $\hbar \to 0$,
the differential operator $X_f^\star$ becomes the vector field $X_f$ as a first-order differential operator,
the quasiprobability distribution becomes the probability distribution,
the noncommutative symplectomorphism in \eqref{U-cl.star}
becomes the time-evolution symplectomorphism in \eqref{U-cl},
and
the $\hbar$-deformed exponential generator in \eqref{Gsol.star}
becomes the exponential generator in \eqref{Gsol}.

Application to the scattering context is straightforward.
The bottom line is that
the S-matrix \textit{is} a noncommutative symplectomorphism on phase space
in the phase space formulation,
which mechanically reduces to a symplectomorphism on phase space
in the commutative limit $\hbar\to0$.
Therefore, it is established that the S-symplectomorphism is the precise classical limit of the S-matrix in a strict sense.

We shall elaborate more on this formalization in a later work \cite{S-symp}.

\section{More on Magnus Series}
\label{MAGPF}

Here, we sketch the proof
of the recursion relation given in \eqref{recursion.magnus}.
We will be brief,
as this is a well-established and well-reviewed result
\cite{magnus1954exponential,eikonal2}.
We will also record
an alternative formula due to Chen and Strichartz \cite{strichartz1987campbell}.

To state the problem, 
we introduce a temporary parameter $\e$
within this appendix.
Consider the exponential generator to a variable intermediate time:
\begin{align}
	\label{mdef}
	\exp(X_{\G(t)})
	\,&=\,
		\Pexp{\,
			\e\mem
			\int_{0}^{t} dt_1\,\,
				X_{H(t_1)}
		}
	\,,\\
	\label{G(t)}
	\G(t)
	\,&=\,
		\sum_{n=1}^\infty\,
			\e^n\, \G_n(t)
	\,.
\end{align}
We set the initial and final times to $\ti = 0$ and $\tf = T$,
which is a totally arbitrary choice.

By taking a time derivative on \eqref{mdef},
one obtains 
an operator equation
\begin{align}
	\label{dmdef}
	\bb{
	\int_0^1 d\eta\,\,
		\mathe^{\eta\hem [X_{\Omega(t)} , \blank]}
			X_{\dot{\Omega}(t)}
	}
			\exp(X_{\G(t)})
	\,=\,
		\e\mem X_{H(t)}\mem 
			\exp(X_{\G(t)})
	\,.
\end{align}
where we have used
a well-known result \cite{rossmann2006lie,hall2013lie}
to evaluate the left-hand side.
Performing the integral in \eqref{dmdef}, it follows that
\begin{align}
	\label{dotG(t)}
    \dot{\G}(t)
    \,=\, 
    \e\,
    \sum_{k=0}^\infty\,
        \frac{B_k}{k!}\,
            (X_{\G(t)})^k\mem
            H(t)
	\,,
\end{align}
where $B_k$ denotes the $k$\textsuperscript{th} Bernoulli number
defined in \eqref{bernoulli}.
Finally, plugging in \eqref{G(t)} to \eqref{dotG(t)} yields \eqref{recursion.magnus}.

Note that the Bernoulli number $B_k$ vanishes for all odd $k$ greater than $1$:
$B_0 = 1$,
$B_1 = -1/2$,
$B_2 = 1/6$,
$B_3 = 0$,
$B_4 = -1/30$,
$B_5 = 0$,
$B_6 = 1/42$,
$B_7 = 0$,
$B_8 = -1/30$,
$\cdots$.
This property plays certain roles
in performing the recursion.

Meanwhile, another useful result known for the Magnus series
is the Chen-Strichartz formula \cite{strichartz1987campbell},
which expands out the nested brackets fully:
\begin{subequations}
\begin{align}
    \label{eq:CSformula}
    \G_n(t)
    \,=\,
    \int_{t_1 > t_2 > \cdots > t_n}\nem d^nt\,\,
    \sum_{\sigma \in \mathrm{S}_n}
        \frac{(-1)^{a_\sigma}}{
            n^2\mem \binom{n-1}{a_\sigma}
        }\,\,
   			\pb{H(t_{\s(1)})}{
   				\pb{H(t_{\s(2)})}{
   					\cdots
   						\pb{H(t_{\s(n-1)})}{
   							H(t_{\s(n)})
   						}
   				}
   			}
    \,.
\end{align}
Here, $\sigma$ runs through permutations of $\{1,2,\cdots,n\}$,
and $a_\sigma$ denotes the number of ascents in the permutation $\sigma$.
Note that this can be equivalently rewritten as
\begin{align}
    \label{eq:CSformula-d}
    \G_n(t)
    \,=\,
    \int_{t_1 > t_2 > \cdots > t_n} d^nt\,\,
    \sum_{\sigma \in \mathrm{S}_n}
        \frac{(-1)^{d_\sigma}}{
            n^2\mem \binom{n-1}{d_\sigma}
        }\,\,
            \{ \cdots \{ \{ H(t_{\s(1)}) , H(t_{\s(2)}) \} , H(t_{\s(3)}) \} \cdots , H(t_{\s(n)}) \}
    \,,
\end{align}
\end{subequations}
where $d_\sigma$ denotes the number of descents in $\sigma$.

\bibliography{references.bib}

\end{document}